\documentclass[conference]{IEEEtran}
\usepackage[left=0.75in,right=0.75in,top=1in,bottom=0.75in]{geometry}
\usepackage{graphicx}
\usepackage[cmex10]{amsmath}
\usepackage{amsfonts}
\usepackage[ruled]{algorithm2e}
\usepackage{todonotes}
\SetKwComment{Comment}{$\triangleright$\ }{}

\newtheorem{lemma}{Lemma}
\newtheorem{definition}{Definition}
\newtheorem{theorem}{Theorem}

\begin{document}
\title{Straggler Resilient Serverless Computing Based on Polar Codes} %

\author{
\IEEEauthorblockN{Burak Bartan and Mert Pilanci}
\IEEEauthorblockA{Department of Electrical Engineering\\
Stanford University\\
Email: {\{bbartan, pilanci\}@stanford.edu}
}}

\maketitle

\begin{abstract}
We propose a serverless computing mechanism for distributed computation based on polar codes. Serverless computing is an emerging cloud based computation model that lets users run their functions on the cloud without provisioning or managing servers. Our proposed approach is a hybrid computing framework that carries out computationally expensive tasks such as linear algebraic operations involving large-scale data using serverless computing and does the rest of the processing locally. We address the limitations and reliability issues of serverless platforms such as straggling workers using coding theory, drawing ideas from recent literature on coded computation. The proposed mechanism uses polar codes to ensure straggler-resilience in a computationally effective manner. We provide extensive evidence showing polar codes outperform other coding methods. We have designed a sequential decoder specifically for polar codes in erasure channels with full-precision input and outputs. In addition, we have extended the proposed method to the matrix multiplication case where both matrices being multiplied are coded. The proposed coded computation scheme is implemented for AWS Lambda. Experiment results are presented where the performance of the proposed coded computation technique is tested in optimization via gradient descent. Finally, we introduce the idea of partial polarization which reduces the computational burden of encoding and decoding at the expense of straggler-resilience. 



\end{abstract}

\section{Introduction}
Computations for large-scale problems on the cloud might require a certain level of expertise for managing servers, and certainly requires configuration of resources ahead of time. It might be challenging to estimate beforehand exactly how much RAM, storage, and time a certain problem requires. For server-based computing on the cloud, users have to manage these configurations which if not done right, costs more money for users. Serverless computing, on the other hand, emerges as a platform simpler to use with much less overhead. Serverless functions offer an alternative to large-scale computing with a lot to offer in terms of cost and speed. 

In this work, we work with a mechanism based on serverless computing that enables users to run their computations in their computers except whenever there is a large-scale matrix multiplication, the computation is done in parallel in a serverless computing platform. The key advantages of this mechanism that may not be available in server-based approaches are:
\begin{itemize}
    \item When serverless functions are not being used (during some local computations in an algorithm), since there are no servers running, users do not pay for the resources they do not use.
    \item The serverless approach has the ease of running the overall algorithm in your own machine with only computationally-heavy parts running on the cloud and returning their outputs to your local machine for further manipulation.
    \item It is possible to request different numbers of serverless functions in different stages of the algorithm. This might be cost-effective when running algorithms with varying computational loads.
\end{itemize}
Along with these advantages, serverless computing comes with certain limitations for resources such as RAM and lifetime. For instance, for AWS Lambda, each function can have a maximum of $3$ GB of RAM and has a lifetime up to $15$ minutes. A mechanism based on serverless computing must take these limitations into account.

One problem that serverless computing platforms and server-based platforms have in common is that a portion of the functions (or workers) might finish their task later than the others (these slower workers are often referred to as stragglers), and this may cause the overall output to be delayed if no precaution is taken. Furthermore, it is not uncommon in serverless computing for a function to return nothing because functions can crash due to the lifetime constraint or for other reasons. The straggler issue could be more problematic in the case of serverless computing than in server-based computing. For instance, if we assume that each function will straggle with a given probability independent of other functions, the probability that there will be at least one straggling function increases as the number of functions increases. Since each function can have a limited amount of memory, large-scale computations usually require many functions to run in parallel. 

To overcome the issue of straggling functions, we use the idea of inserting redundancy to computations using codes. This idea has been explored in the literature by many works such as \cite{Lee2018}, \cite{baharav2018prodcodes}, \cite{yu2017polycode} for dealing with stragglers encountered in distributed computation. Codes not only help speed up computations by making it possible to compute the desired output without waiting for the outputs of the straggling workers, but also provide resilience against crashes and timeouts, leading to a more reliable mechanism.

\newgeometry{top=0.75in,bottom=0.75in,right=0.75in,left=0.75in}

A straggler-resilient scheme designed for a serverless computing platform needs to be scalable as we wish to allow the number of functions to vary greatly. Two things become particularly important if the number of workers is as high as hundreds or thousands. The first one is that encoding and decoding must be low complexity. The second one is that one must be careful with the numerical round-off errors if the inputs are not from a finite field, but instead are full-precision real numbers. To clarify this point, when the inputs are real-valued, encoding and decoding operations introduce round-off errors. Polar codes show superiority over many codes in terms of both of these aspects. They have low encoding and decoding complexity and both encoding and decoding involve a small number of subtraction and addition operations (no multiplications involved in encoding or decoding). In addition, they achieve capacity, and the importance of that fact in coded computation is that the number of worker outputs needed for decoding is asymptotically optimal.


\subsection{An Important Difference Between Server-Based and Serverless Computing}
We should note one important difference between using serverless and server-based computing that helps highlight the usefulness of polar codes. In server-based computing, one needs to use much fewer machines than the number of functions one would need in serverless computing to achieve the same amount of computing. The reason for this is the limited resources each function can have in serverless computing. Hence, the number of functions is usually expected to be high, depending on the scale of the computation. This is an important design criterion that needs to be addressed as it necessitates efficient encoding and decoding algorithms for the code we are using. For instance, for a distributed server-based system with $N=8$ machines, using maximum distance separable (MDS) codes with decoding complexity as high as $O(N^3)$ could still be acceptable. This becomes unacceptable when we switch to a serverless system as the number of functions that achieve the same amount of computations could go up as high as a few hundreds. In this case, it becomes necessary to use codes with fast decoders such as polar codes.

Even though polar codes do not have the MDS code properties, as the block-length of the code increases (i.e. more functions are used), the performance gap between polar codes and MDS codes closes. We further discuss this point in the Numerical Results section.

\subsection{Related Work}
Coded matrix multiplication has been introduced in \cite{Lee2018} for speeding up distributed matrix-vector multiplication in server-based computing platforms. \cite{Lee2018} has shown that it is possible to speed up distributed matrix multiplication by using codes (MDS codes in particular) to avoid waiting for the slowest nodes. MDS codes however have the disadvantage of having high encoding and decoding complexity, which could be restricting in setups with large number of workers. \cite{baharav2018prodcodes} attacks at this problem presenting a coded computation scheme based on $d$-dimensional product codes. \cite{yu2017polycode} presents a scheme referred to as polynomial codes for coded matrix multiplication with input matrices from a large finite field. This approach might require quantization for real-valued inputs which could introduce additional numerical issues. \cite{dutta2018optimalrec} and \cite{yu2018straggler} are other works investigating coded matrix multiplication and provide analysis on the optimal number of worker outputs required. In addition to the coding theoretic approaches, \cite{gupta2018oversketch} offers an \textit{approximate} straggler-resilient matrix multiplication scheme where the ideas of sketching and straggler-resilient distributed matrix multiplication are brought together.

Using Luby Transform (LT) codes, a type of rateless fountain codes, in coded computation has been recently proposed in \cite{severinson2018lt} and \cite{mallick2018rateless}. The proposed scheme in \cite{mallick2018rateless} divides the overall task into smaller tasks (each task is a multiplication of a row of $A$ with $x$) for better load-balancing. The work \cite{severinson2018lt} proposes the use of inactivation decoding and the work \cite{mallick2018rateless} uses peeling decoder. Peeling decoder has a computational complexity of $O(N\log N)$ (same complexity as the decoder we propose), however its performance is not satisfactory if the number of inputs symbols is not very large. Inactivation decoder performs better than the peeling decoder, however, it is not as fast as the peeling decoder.

There has been some recent work on serverless computing for machine learning training. In \cite{carreira2018caseserverless}, serverless machine learning training is discussed in detail, and challenges and possible solutions for designing serverless machine training are provided. Authors in \cite{feng2018serverlesstraining} consider an architecture with multiple master nodes and state that for small neural network models, serverless computing helps speed up hyperparameter tuning. Similarly, the work in \cite{wang2019serverlesslearning} shows via experiments that their prototype on AWS Lambda can reduce model training time greatly. 
\subsection{Main Contributions}
This work investigates the use of serverless computing for large-scale computations with a solution to the issue of stragglers based on polar codes. We identify that polar codes are a natural choice for coded computation with serverless computing due to their low complexity encoding and decoding. Furthermore, we propose a sequential decoder for polar codes designed to work with full-precision data for erasure channels. The proposed decoder is different from successive cancellation (SC) decoder given in \cite{polar2009arikan}, which is for discrete data and is based on estimation using likelihoods. We also analytically justify the use of the polar code kernel that we used in this work when the inputs and outputs are not from a finite field but are full-precision real-valued data. Next, we discuss the polarization of distributions of worker run times and provide some numerical results to that end. This provides some important insights towards understanding polarization phenomenon in a setting other than communication. 


Given the work of \cite{Lee2018}, it is not very surprising that any linear code like polar codes could also be used in coded computation. The main novelty of this work lies in the design of a mechanism based on serverless computing that is easy to use and can handle large-scale data. We adopt coded computation as the tool to make the mechanism more resilient and reliable, and polar codes are our choice of code to ensure scalability.


Differently from other erasure codes, polar codes offer a very interesting interpretation in terms of run times of machines due to the polarization phenomenon. By bringing together a number of machines in polar code construction, we achieve virtually transformed machines with polarized run times. To clarify, the polar coding transformation polarizes the run times of the workers, that is, as a result of the transformation, we will obtain faster and slower transformed machines. This interpretation is equivalent to channels polarizing when polar codes are used for channel coding in communications \cite{polar2009arikan}. If we set the rate to $1-\epsilon$, then we choose the best $N(1-\epsilon)$ machines out of the $N$ machines as the data machines whereas the remaining machines are frozen (i.e. we send in zero matrices for the slow machines).


\section{Coded Computation with Polar Coding}
In this section we describe our proposed polar coding method for matrix-vector multiplication. Later in coded matrix multiplication section, we talk about the extension of this method to matrix multiplication (both matrices are encoded instead of one of them). We would like to note that the method described in this section is applicable for both server-based and serverless computing platforms. Hence, we use the terms \textit{worker}, \textit{serverless function}, \textit{compute node} to mean the same thing, which is a single computing block of whatever distributed platform we are using. The term master node refers to the machine that we aggregete the results in.

\subsection{System Model}
\begin{figure}[htbp]
  \centering
  \includegraphics[width=0.48\textwidth]{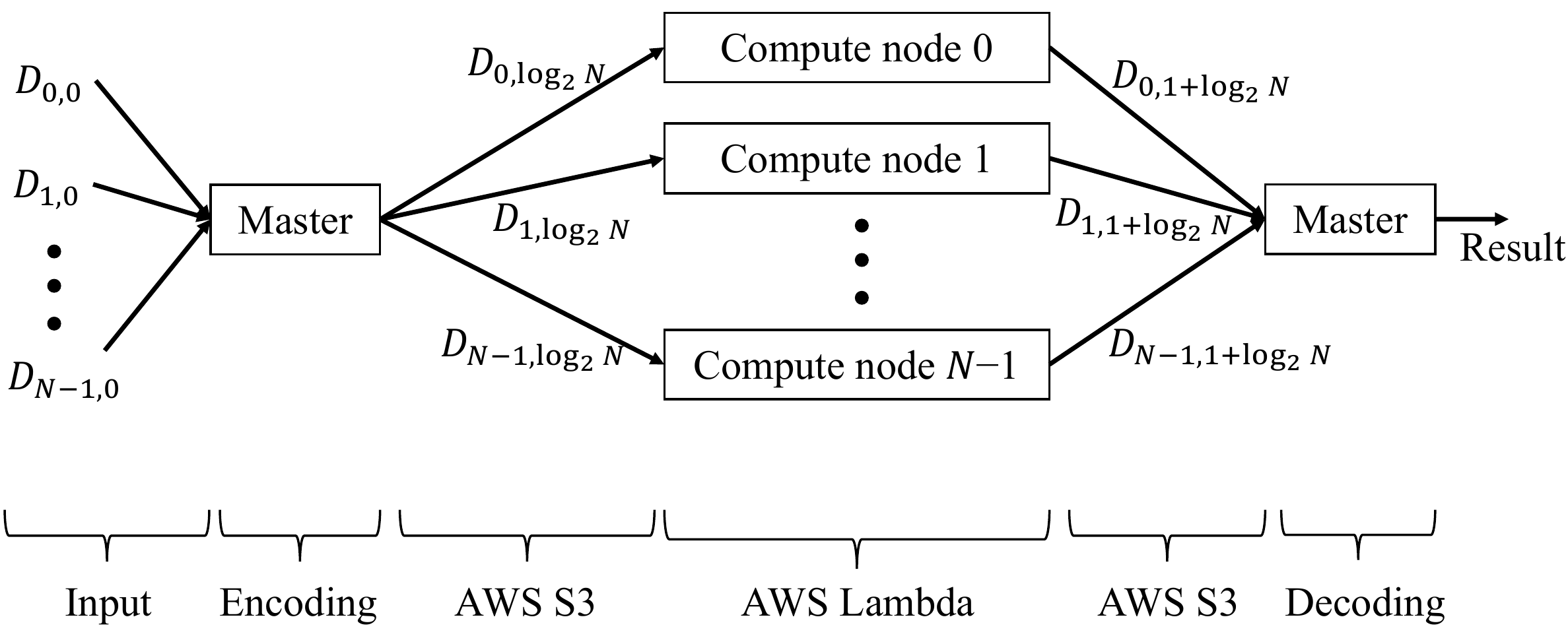}
  \caption{System model.}
  \label{architecture}
\end{figure}

We are interested in speeding up the computation of $A\times x$ where $A\in \mathbb{R}^{m\times n}$ and $x \in \mathbb{R}^{n\times r}$ using the model in Figure \ref{architecture}. The master node encodes the $A$ matrix and communicates the encoded matrix chunks to workers. Compute nodes then perform their assigned matrix multiplication. After compute nodes start performing their assigned tasks, the master node starts listening for compute node outputs. Whenever a decodable set of outputs is detected, the master node downloads the available node outputs and performs decoding, using the sequential decoder described in subsection II.E.

Note that this approach assumes $x$ fits in the memory of a compute node. In coded computation literature, this is referred to as large-scale matrix-vector multiplication, different from matrix-matrix multiplication where the second matrix is also partitioned. We discuss that case later in the text. 

\subsection{Code Construction}
\begin{figure}[htbp]
  \centering
  \includegraphics[width=0.3\textwidth]{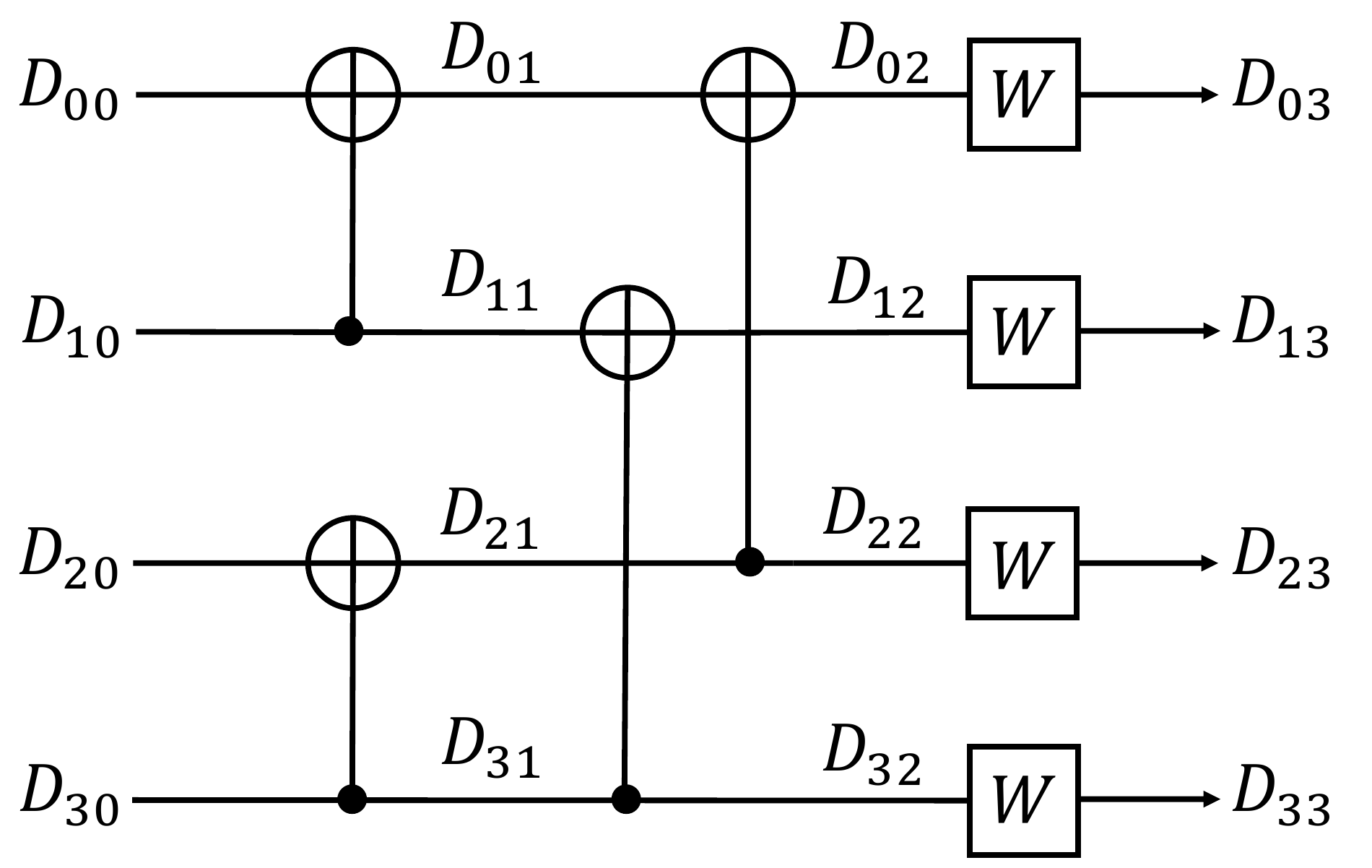}
  \caption{Polar coding for $N=4$.}
  \label{circuit_fig}
\end{figure}

Consider the polar coding representation in Figure \ref{circuit_fig} for $N=4$ nodes. The channels $W$ shown in Figure \ref{circuit_fig} are always erasure channels in this work because of the assumption that if the output of a compute node is available, then it is assumed to be correct. If the output of a node is not available (straggler node), then it is considered an erasure. 

Let $D_{ij}$ denote the data for the node $i$ in the $j$'th level \footnotemark. Further, let us denote the erasure probability of each channel by $\epsilon$, and assume that the channels are independent. Using the notation from \cite{polar2009arikan}, the coding scheme in Figure \ref{circuit_fig} does the following channel transformation: $W^4 \to (W_4^{(0)},...,W_4^{(3)})$ where the transformed channels are defined as:
\begin{align}
    W_4^{(0)}&: D_{00} \to (D_{03},...,D_{33}) \nonumber\\
    W_4^{(1)}&: D_{10} \to (D_{03},...,D_{33}, D_{00}) \nonumber\\
    W_4^{(2)}&: D_{20} \to (D_{03},...,D_{33}, D_{00}, D_{10}) \nonumber\\
    W_4^{(3)}&: D_{30} \to (D_{03},...,D_{33}, D_{00}, D_{10}, D_{20})
\end{align}

\footnotetext{Note that during encoding, the node values $D_{ij}$ represent the data before it is multiplied by $x$, and during decoding, $D_{ij}$s represent the data after the multiplication by $x$.}

\begin{figure}[htbp]
  \centering
  \includegraphics[width=0.23\textwidth]{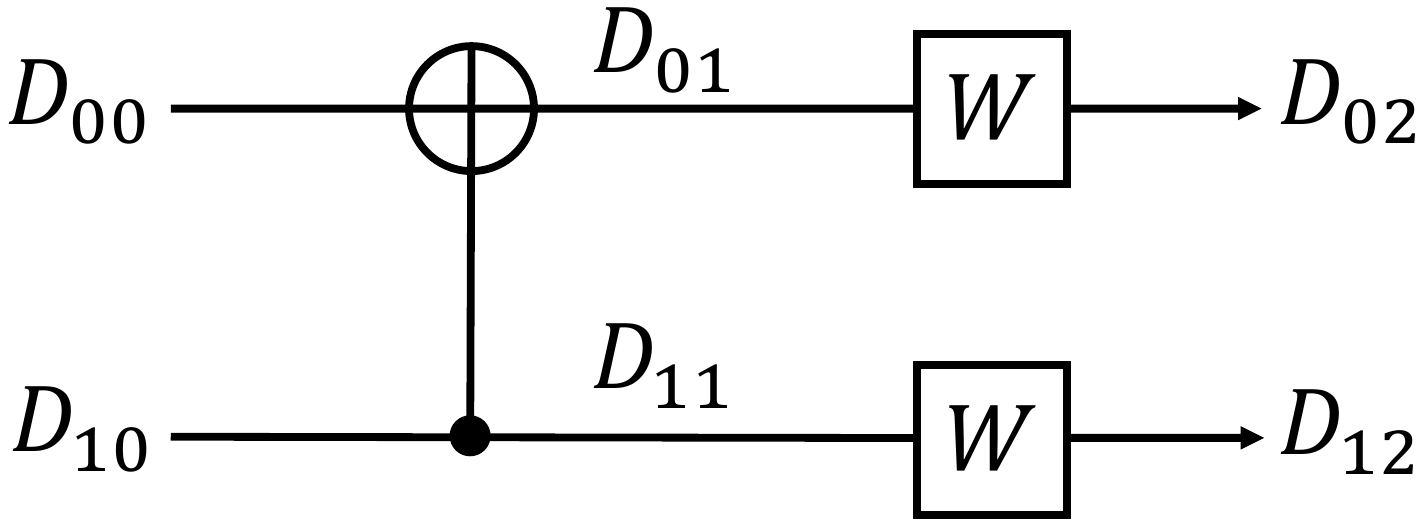}
  \caption{Polar coding for $N=2$.}
  \label{circuit_fig_N_2}
\end{figure}

The transformation for $N=4$ in fact consists of two levels of two $N=2$ blocks, where an $N=2$ block does the transformation $W^2 \to (W_2^{(0)}, W_2^{(1)})$, which is shown in Figure \ref{circuit_fig_N_2}. We can now compute the erasure probabilities for the transformed channels. Consider the block for $N=2$ in Figure \ref{circuit_fig_N_2}. We have $Pr(D_{00} \text{ is erased}) = 1-(1-\epsilon)^2$, $Pr(D_{01} \text{ is erased} | D_{00}) = \epsilon^2$.

To compute the erasure probabilities for $N=4$, note that $D_{01}, D_{21}, D_{03}, D_{23}$ form an $N=2$ block. Similarly, $D_{11}, D_{31}, D_{13}, D_{33}$ form another $N=2$ block. It follows that the erasure probabilities for the level $1$ are: $Pr(D_{01} \text{ is erased}) = Pr(D_{11} \text{ is erased}) = 1-(1-\epsilon)^2$, and $Pr(D_{21} \text{ is erased} |D_{01}) = Pr(D_{31} \text{ is erased} | D_{11}) = \epsilon^2$. Now, we can compute the erasure probabilities for the input level:
\begin{align} \label{erasure_probs}
    &Pr(D_{00} \text{ is erased}) = 1 - (1-\epsilon)^4 \nonumber \\
    &Pr(D_{10} \text{ is erased} | D_{00}) = (1-(1-\epsilon)^2)^2 \nonumber\\
    &Pr(D_{20} \text{ is erased} | D_{00}, D_{10}) = 1 - (1-\epsilon^2)^2 \nonumber\\
    &Pr(D_{30} \text{ is erased} | D_{00}, D_{10}, D_{20}) = \epsilon^4
\end{align}
It is straightforward to repeat this procedure to find the erasure probabilities for the transformed channels in the cases of larger $N$.

Knowing the erasure probabilities of channels, we now select the best channels for data, and freeze the rest (i.e., set to zero). The indices of data and frozen channels are made known to both encoding and decoding. Note that by 'channel', we mean the transformed virtual channels $W_N^{(i)}$, not $W$. We refer the readers to \cite{polar2009arikan} for an in-depth analysis and discussion on polar codes.

A possible disadvantage of the proposed scheme is that the number of compute nodes has to be a power of $2$. This could pose a problem if the number of workers we are interested in utilizing is not a power of $2$. We identify two possible solutions to overcome this issue. The first one is to use a kernel of different size instead of $2$. The second solution is to have multiple code constructions. For instance, let us assume that we wish to use $20$ compute nodes and we choose $\epsilon=\frac{1}{4}$. We can divide the overall task into $2$ constructions. We would compute the first task using $16$ compute nodes ($12$ data, $4$ frozen channels), and the second task using $4$ compute nodes ($3$ data, $1$ frozen channels), where the first task is to multiply the first $\frac{4}{5} m$ rows of $A$ with $x$, and the second task is to multiply the last $\frac{1}{5} m$ rows of $A$ with $x$. This way we obtain two straggler-resilient tasks and we use all $20$ workers.

\subsection{Encoding Algorithm}
After computing the erasure probabilities for the transformed channels, We choose the $d = N (1-\epsilon)$ channels with the lowest erasure probabilities as data channels. The remaining $N\epsilon$ channels are frozen. The inputs for the data channels are the row partitions of matrix $A$. That is, we partition $A$ such that each matrix chunk has $\frac{m}{d}$ rows: $A=[A_1^T, A_2^T, \dots A_d^T]^T$ where $A_i \in \mathbb{R}^{\frac{m}{d}\times n}$. For example, for $N=4$ and $\epsilon = 0.5$, the erasure probabilities of the transform channels are calculated using \eqref{erasure_probs} to be $[0.938, 0.563, 0.438, 0.063]$. It follows that we freeze the first two channels, and the last two channels are the data channels. This means that in Figure \ref{circuit_fig}, we set $D_{00}=D_{10}=0^{\frac{m}{2}\times n}$ and $D_{20}=A_1$, $D_{30}=A_2$.

We encode the input using the structure given for $N=4$ in Figure \ref{circuit_fig}. Note that instead of the XOR operation like in polar coding for communication, what we have in this work is addition over real numbers. This structure can be easily formed for higher $N$ values. For instance, for $N=8$, we simply connect the circuit in Figure \ref{circuit_fig} with its duplicate placed below it. The way encoding is done is that we compute the data for each level from left to right starting from the leftmost level. There are $\log_2 N + 1$ levels in total and $N$ nodes in every level. Hence, the encoding complexity is $O(N\log N)$.

\subsection{Channel}
Channels in the communication context correspond to compute nodes in the coded computation framework. The $i$'th node computes the matrix multiplication $D_{i,\log_2{N}} \times x$. When a decodable set of outputs is detected, the master node moves on to decoding the available outputs. At that time, all the delayed nodes (stragglers) are considered to be erased.


\subsection{Decoding Algorithm}
The decoding algorithm, whose pseudo-code is given in Alg. \ref{decoding_alg}, does not require quantization of data as it is tailored to work with full-precision data. Note that the notation $I_{D_{ij}} \in \{0,1\}$ indicates whether the data for node $i$ in level $j$ (namely, $D_{ij}$) is known. In the first part of the decoding algorithm, we continuously check for decodability as compute node outputs become available. When the available outputs become decodable, we move on to the second part. In the second part, decoding takes place given the available downloaded compute node outputs. The second part makes calls to the function \textit{decodeRecursive}, given in Alg. \ref{decode_rec}.

\DontPrintSemicolon

\begin{algorithm}
 \KwIn{Indices of the frozen channels}
 \KwResult{$y = A \times x$ \Comment*[r]{Part I}}
 
 Initialize $I_{D_{:,0}} = [I_{D_{0,0}}, I_{D_{1,0}}, \dots, I_{D_{N-1,0}}] = [0,\dots,0]$ \;
 \While{$I_{D_{:,0}}$ not decodable}{
 update $I_{D_{:,0}}$\;
 checkDecodability($I_{D_{:,0}}$)\footnotemark
 }
 Initialize an empty list $y$  \Comment*[r]{Part II}
 \For{$i \gets 0$ \textbf{to} $N-1$} {
   $D_{i,0}$ = decodeRecursive($i$, $0$)\;
   \If{node $i$ is a data node} {
     $y = [y; D_{i,0}]$\;
   }
   \Comment*[r]{forward prop}
   \vspace{-0.3cm}
   \If{$i \bmod 2 = 1$ } {
     \For{$j \gets 0$ \textbf{to} $\log_2{N}$} {
       \For{$l \gets 0$ \textbf{to} $i$}{
         compute $D_{lj}$ if unknown 
       }
     }
   }
  }
 \caption{Decoding algorithm.}
 \label{decoding_alg}
\end{algorithm}

\footnotetext{checkDecodability is a function that checks whether it is possible to decode a given sequence of indicators (identifying availability of outputs). This works the same way as part II of Algorithm \ref{decoding_alg}, but differs in that it does not deal with data, but instead binary indicators.}

\begin{algorithm}
 \KwIn{Node $i \in [0,N-1]$, level $j \in [0,\log_2{N}]$}
 \KwResult{$I_{D_{ij}}$ and modifies $D$}
 
  \lIf{$j = \log_2{N}$} { 
     return $I_{D_{i,\log_2{N}}}$ \Comment*[r]{base case 1}
   }
   \vspace{-0.4cm}
  \lIf{$I_{D_{ij}} = 1$} {
     return $1$ \Comment*[r]{base case 2}
   }
   \vspace{-0.4cm}
  $I_{D_{i,(j+1)}} = $ decodeRecursive($i, j+1$)\;
  $I_{D_{pair(i),(j+1)}} = $ decodeRecursive($pair(i), j+1$)\;
 
  \uIf{$i$ is upper node}{
    \If{$I_{D_{i,(j+1)}} \text{ AND } I_{D_{pair(i),(j+1)}} = 1$}{
      compute $D_{ij}$\;
      return 1
    }
  }
  \Else{
    \If{$I_{D_{i,(j+1)}} \text{ OR } I_{D_{pair(i),(j+1)}}$ = 1}{
      compute $D_{ij}$\;
      return 1
    }
  }
 return 0
 \caption{decodeRecursive($i$, $j$)}
 \label{decode_rec}
\end{algorithm}

\section{Analysis}
\subsection{Polar Coding on Full-Precision Data}
Note that both encoding and decoding algorithms operate only on $N=2$ blocks which correspond to the kernel $F_2=\bigl[\begin{smallmatrix}1 & 1\\ 0 & 1\end{smallmatrix}\bigr]$. Theorem \ref{thm_decoder_kernel} justifies the use of $F_2$ for real-valued data among all possible $2\times 2$ kernels. We first give a definition for a \textit{polarizing kernel}, and then give a lemma which will later be used in proving Theorem \ref{thm_decoder_kernel}.
\begin{definition}[Polarizing kernel] \label{polarizing_kernel_def}
Let $f$ be a function satisfying the linearity property $f(au_1+bu_2) = af(u_1)+bf(u_2)$ where $a,b \in \mathbb{R}$ and assume that there is an algorithm to compute $f$ that takes a certain amount of time to run with its run time distributed randomly. Let $K$ denote a $2\times 2$ kernel and $\bigl[\begin{smallmatrix} v_1 \\ v_2 \end{smallmatrix}\bigr] = K \times \bigl[\begin{smallmatrix} u_1 \\ u_2 \end{smallmatrix}\bigr]$. Assume that we input $v_1$ and $v_2$ to two i.i.d. instances of the same algorithm for $f$. Further, let $T_1$, $T_2$ be random variables denoting the run times for computing $f(v_1)$, $f(v_2)$, respectively. We are interested in computing $f(u_1)$, $f(u_2)$ in this order. If the time required to compute $f(u_1)$ is $\max(T_1,T_2)$ and the time required to compute $f(u_2)$ given the value of $f(u_1)$ is $\min(T_1,T_2)$, then we say $K$ is a polarizing kernel.
\end{definition}

\begin{lemma} \label{polarizing_kernel_lemmma}
A kernel $K \in \mathbb{R}^{2 \times 2}$ is a polarizing kernel if and only if the following conditions are both satisfied: 1) Both elements in the second column of $K$ are nonzero, 2) $K$ is invertible.
\end{lemma}
\begin{IEEEproof}
We first prove that if $K$ is a polarizing kernel, then it satisfies both of the given conditions. Let us assume an $f$ function satisfying the linearity property given in Definition \ref{polarizing_kernel_def}. Since $f$ holds the linearity property, we can write
\begin{align}
    \left[\begin{matrix} f(v_1) \\ f(v_2) \end{matrix}\right] = \left[\begin{matrix}K_{11} & K_{12}\\ K_{21} & K_{22} \end{matrix}\right] \times \left[\begin{matrix} f(u_1) \\ f(u_2) \end{matrix}\right].
\end{align} 
Computing $f(u_2)$ given the value of $f(u_1)$ in time $\min(T_1,T_2)$ means that $f(u_2)$ can be computed using $f(u_1)$ and either one of $f(v_1)$, $f(v_2)$ (whichever is computed earlier). This implies that we must be able to recover $f(u_2)$ using one of the following two equations
\begin{align} 
    K_{12} \times f(u_2) &= f(v_1) - K_{11}f(u_1) \label{lemma_eq_1} \\
    K_{22} \times f(u_2) &= f(v_2) - K_{21}f(u_1). \label{lemma_eq_2}
\end{align}
We use \eqref{lemma_eq_1} if $f(v_1)$ is known, and \eqref{lemma_eq_2} if $f(v_2)$ is known. This implies that both $K_{12}$ and $K_{22}$ need to be nonzero. Furthermore, to be able to compute $f(u_1)$ in time $\max(T_1, T_2)$ means that it is possible to find $f(u_1)$ using both $f(v_1)$ and $f(v_2)$ (note that we do not assume we know the value of $f(u_2)$). There are two scenarios where this is possible: Either at least one row of $K$ must have its first element as nonzero and its second element as zero, or $K$ must be invertible. Since we already found out that $K_{12}$ and $K_{22}$ are both nonzero, we are left with one scenario, which is that $K$ must be invertible.

We proceed to prove the other direction of the 'if and only if' statement, which states that if a kernel $K \in \mathbb{R}^{2 \times 2}$ satisfies the given two conditions, then it is a polarizing kernel. We start by assuming an invertible $K \in \mathbb{R}^{2 \times 2}$ with both elements in its second column nonzero. Since $K$ is invertible, we can uniquely determine $f(u_1)$ when both $f(v_1)$ and $f(v_2)$ are available, which occurs at time $\max(T_1,T_2)$. Furthermore, assume we know the value of $f(u_1)$. At time $\min(T_1,T_2)$, we will also know one of $f(v_1)$, $f(v_2)$, whichever is computed earlier. Knowing $f(u_1)$, and any one of $f(v_1)$, $f(v_2)$, we can determine $f(u_2)$ using the suitable one of the equations \eqref{lemma_eq_1}, \eqref{lemma_eq_2} because $K_{12}$ and $K_{22}$ are both assumed to be nonzero. Hence this completes the proof that a kernel $K$ satisfying the given two conditions is a polarizing kernel.
\end{IEEEproof}


\begin{theorem} \label{thm_decoder_kernel}
 Of all possible $2\times 2$ polarizing kernels, the kernel $F_2=\bigl[\begin{smallmatrix}1 & 1\\ 0 & 1\end{smallmatrix}\bigr]$ results in the fewest number of computations for encoding real-valued data.
\end{theorem}
\begin{IEEEproof}
By Lemma 1, we know that for $K$ to be a polarizing kernel, it must be invertible and must have both $K_{12}$ and $K_{22}$ as nonzero. For a $2\times 2$ matrix to be invertible with both second column elements as nonzero, at least one of the elements in the first column must also be nonzero. We now know that $K_{12}$, $K_{22}$ and at least one of $K_{11}$, $K_{21}$ must be nonzero in a polarizing kernel $K$. It is easy to see that having all four elements of $K$ as nonzero leads to more computations than having only three elements of $K$ as nonzero. Hence, we must choose either $K_{11}$ or $K_{21}$ to be zero (it does not matter which one). It is possible to avoid any multiplications by selecting the nonzero elements of $K$ as ones. Hence, both $K=\bigl[\begin{smallmatrix}1 & 1\\ 0 & 1\end{smallmatrix}\bigr]$ and $K=\bigl[\begin{smallmatrix}0 & 1\\ 1 & 1\end{smallmatrix}\bigr]$ are polarizing kernels and lead to the same amount of computations, which is a single addition. This amount of computations is the minimum possible as otherwise $K$ will not satisfy the condition that a polarizing kernel must have at least $3$ nonzero elements.
\end{IEEEproof}

\section{Discussion on Fast Decoders for MDS Codes}

Most works in coded computation literature employ MDS codes as their way of inserting redundancy into computations. Decoding in the case of Reed-Solomon codes and full-precision data requires solving a linear system. In the case where one wishes to work with full-precision data, the decoding becomes expensive (using Gaussian elimination it takes $O(n^3)$ operations) and in addition gets unstable for systems with high number of workers since the linear system to be solved is a Vandermonde based linear system.

There are many works that restrict their schemes to working with values from a finite field of some size $q$. In that case, it is possible to use fast decoding algorithms which are based on fast algorithms for polynomial interpolation. One such decoding algorithm is given in \cite{soro2009rs_decoder}, which provides $O(n\log n)$ encoding and decoding algorithms for Reed-Solomon erasure codes based on Fermat Number Transform (FNT). The complexity for the encoder is the same as taking a single FNT transform and for the decoder, it is equal to taking $8$ FNT transforms. To compare with polar codes, polar codes require $n\log n$ operations for both encoding and decoding. Another work where a fast erasure decoder for Reed-Solomon codes is presented is \cite{didier2009rs_decoder} which presents a decoder that works in time $O(n\log^2n)$.



There exist many other fast decoding algorithms for RS codes with complexities as low as $O(n\log n)$. However, the decoding process in these algorithms usually requires taking a fast transform (e.g. FNT) many times and are limited to finite fields. Often these fast decoding algorithms have big hidden constants in their complexity and hence quadratic time decoding algorithms are sometimes preferred over them. Polar codes, on the other hand, provide very straightforward and computationally inexpensive encoding and decoding algorithms. One of our contributions is the design of an efficient decoder for polar codes tailored for the erasure channel that can decode full-precision data.

\section{Partial Construction}
In this section, we introduce a new idea that we refer to as \textit{partial construction}. Let us assume we are interested in computing $Ax$ where we only encode $A$ and not $x$. If the data matrix $A$ is very large, it might be challenging to encode it. A way around having to encode a large $A$ is to consider partial code constructions, that is, for $A=[A_1^T, ..., A_p^T]^T$, we encode each submatrix $A_i$ separately. It follows that decoding the outputs of the construction for the submatrix $A_i$ will give us $A_ix$. This results in a weaker straggler resilience, however, we get a trade-off between the computational load of encoding and straggler resilience. Partial construction also decreases the computations required for decoding since instead of decoding a code with $N$ outputs (of complexity $O(N\log N)$), now we need to decode $p$ codes with $\frac{N}{p}$ outputs, which is of total complexity $O(N\log (\frac{N}{p}))$.

In addition, partial construction makes it possible to parallel compute both encoding and decoding. Each code construction can be encoded and decoded independently from the rest of the constructions.

Partial construction idea can also be applied to coded computation schemes based on other codes. For instance, one scenario where this idea is useful is when one is interested in using RS codes with full-precision data. Given that for large $N$ values, using RS codes with full-precision data becomes impossible, one can construct many smaller size codes. When the code size is small enough, solving a Vandermonde-based linear system can be painlessly done.

\subsection{In-Memory Encoding}
Another benefit of the partial construction idea is that for constructions of sizes small enough, encoding can be performed in the memory of the workers after reading the necessary data. This results in a straggler-resilient scheme without doing any pre-computing to encode the entire dataset. In-memory encoding could be useful for problems where the data matrix $A$ also is also changing over time because it might be too expensive to encode the entire dataset $A$ every time it changes.

\section{Coded Matrix Multiplication}
In this section, we provide an extension to our proposed method to accommodate coding of both $A$ and $B$ for computing the matrix multiplication $AB$ (instead of coding only $A$). This can be thought of as a two-dimensional extension of our method. Let us partition $A$ and $B$ as follows:
\begin{align}
    A = \left[\begin{matrix} A_1 \\ \vdots \\ A_{d_1} \end{matrix}\right], B = \left[\begin{matrix}B_1 & \hdots & B_{d_2} \end{matrix}\right].
\end{align} 
Let us denote zero matrix padded version of $A$ by $\tilde{A}=[\tilde{A}_1^T, ..., \tilde{A}_{N_1}^T]^T$ such that $\tilde{A}_i = 0$ if $i$ is a frozen channel index and $\tilde{A}_i = A_j$ if $i$ is a data channel index with $j$ the appropriate index. Similarly, we define $\tilde{B}=[\tilde{B}_1, ..., \tilde{B}_{N_2}]$ such that $\tilde{B}_i = 0$ if $i$ is a frozen channel index and $\tilde{B}_i =B_j$ if $i$ is a data channel index with $j$ the appropriate index. Encoding on $\tilde{A}$ can be represented as $G_{N_1} \tilde{A}$ where $G_{N_1}$ is the $N_1$ dimensional generator matrix and acts on submatrices $\tilde{A}_i$. Similarly, encoding on $\tilde{B}$ would be $\tilde{B} G_{N_2}$.

Encoding $A$ and $B$ gives us  $N_1$ submatrices $(G_{N_1} \tilde{A})_i$ due to $A$ and $N_2$ submatrices $(\tilde{B} G_{N_2})_j$ due to $B$. Then, we matrix-multiply the encoded matrices using a total of $N_1 N_2$ workers with the $(i,j)$'th worker computing the matrix multiplication $(G_{N_1} \tilde{A})_i (\tilde{B} G_{N_2})_j$. More explicitly, the worker outputs will be of the form:
\begin{align}\label{definition_of_P}
    P = \left[\begin{matrix} (G_{N_1} \tilde{A})_1 (\tilde{B} G_{N_2})_1 & \hdots & (G_{N_1} \tilde{A})_1 (\tilde{B} G_{N_2})_{N_2} \\
    \vdots & \ddots & \\ 
    (G_{N_1} \tilde{A})_{N_1} (\tilde{B} G_{N_2})_1 & \hdots & (G_{N_1} \tilde{A})_{N_1} (\tilde{B} G_{N_2})_{N_2} \end{matrix}\right]
\end{align}
Note that for fixed $j$, the worker outputs are:
\begin{align}
    \left[\begin{matrix} (G_{N_1} \tilde{A})_1 (\tilde{B} G_{N_2})_j  \\ \vdots \\ (G_{N_1} \tilde{A})_{N_1} (\tilde{B} G_{N_2})_j \end{matrix}\right]
\end{align}
For fixed $j$, the outputs are linear in $(\tilde{B} G_{N_2})_j$, hence, it is possible to decode these outputs using the decoder we have for the 1D case. Similarly, for fixed $i$, the outputs are:
\begin{align}
    \left[\begin{matrix} (G_{N_1} \tilde{A})_i (\tilde{B} G_{N_2})_1 & \hdots & (G_{N_1} \tilde{A})_i (\tilde{B} G_{N_2})_{N_2} \end{matrix}\right]
\end{align}
For fixed $i$, the outputs are linear in $(G_{N_1} \tilde{A})_i$. It follows that the 1D decoding algorithm can be used for decoding the outputs. Based on these observations, we designed a polar decoder for the 2D case whose pseudo-code is given in Alg. \ref{2d_decoder}. The 2D decoding algorithm makes calls to the 1D encoding and decoding algorithms many times to fill in the missing entries of the encoded matrix $P$ defined in \eqref{definition_of_P}. When all missing entries of $P$ are computed, first all rows and then all columns of $P$ are decoded and finally, the frozen entries are removed to obtain the multiplication $AB$.

\begin{algorithm}
 \KwIn{the worker output matrix $P$}
 \KwResult{$y = A\times B$}
 
 \While{$P$ has missing entries}{
 \Comment*[r]{loop over rows}
  \For{$i \gets 0$ \textbf{to} $N_1-1$ } {
    \If{$P[i,:]$ has missing entries and is decodable}{
     decode $P[i,:]$ using Alg. \ref{decoding_alg} \\
     forward prop to fill in $P[i,:]$ 
    }
  }
  \Comment*[r]{loop over columns}
  \For{$j \gets 0$ \textbf{to} $N_2-1$} {
    \If{$P[:,j]$ has missing entries and is decodable}{
     decode $P[:,j]$ using Alg. \ref{decoding_alg} \\
     forward prop to fill in $P[:,j]$ 
    }
  }
 }
 decode all rows and then all columns of $P$ \\
 return entries of $P$ (ignoring the frozen entries)
 \caption{2D decoding algorithm for matrix multiplication.}
 \label{2d_decoder}
\end{algorithm}

\section{Numerical Results}
\subsection{Encoding and Decoding Speed Comparison}
\begin{figure}[htb]
\begin{minipage}[b]{0.48\linewidth}
  \centering
  \centerline{\includegraphics[width=4.5cm]{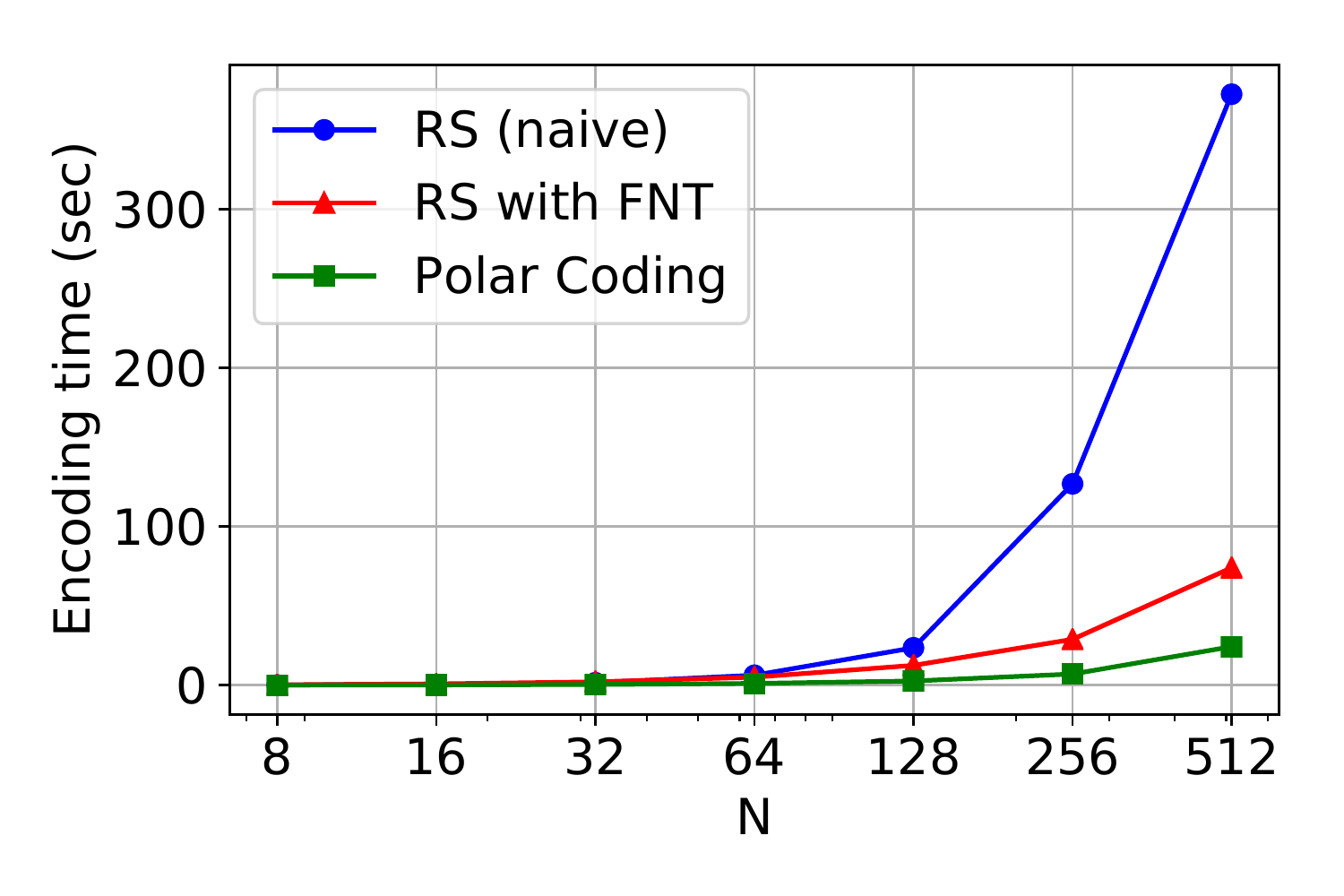}}
  \centerline{(a) Encoding}\medskip
\end{minipage}
\begin{minipage}[b]{0.48\linewidth}
  \centering
  \centerline{\includegraphics[width=4.5cm]{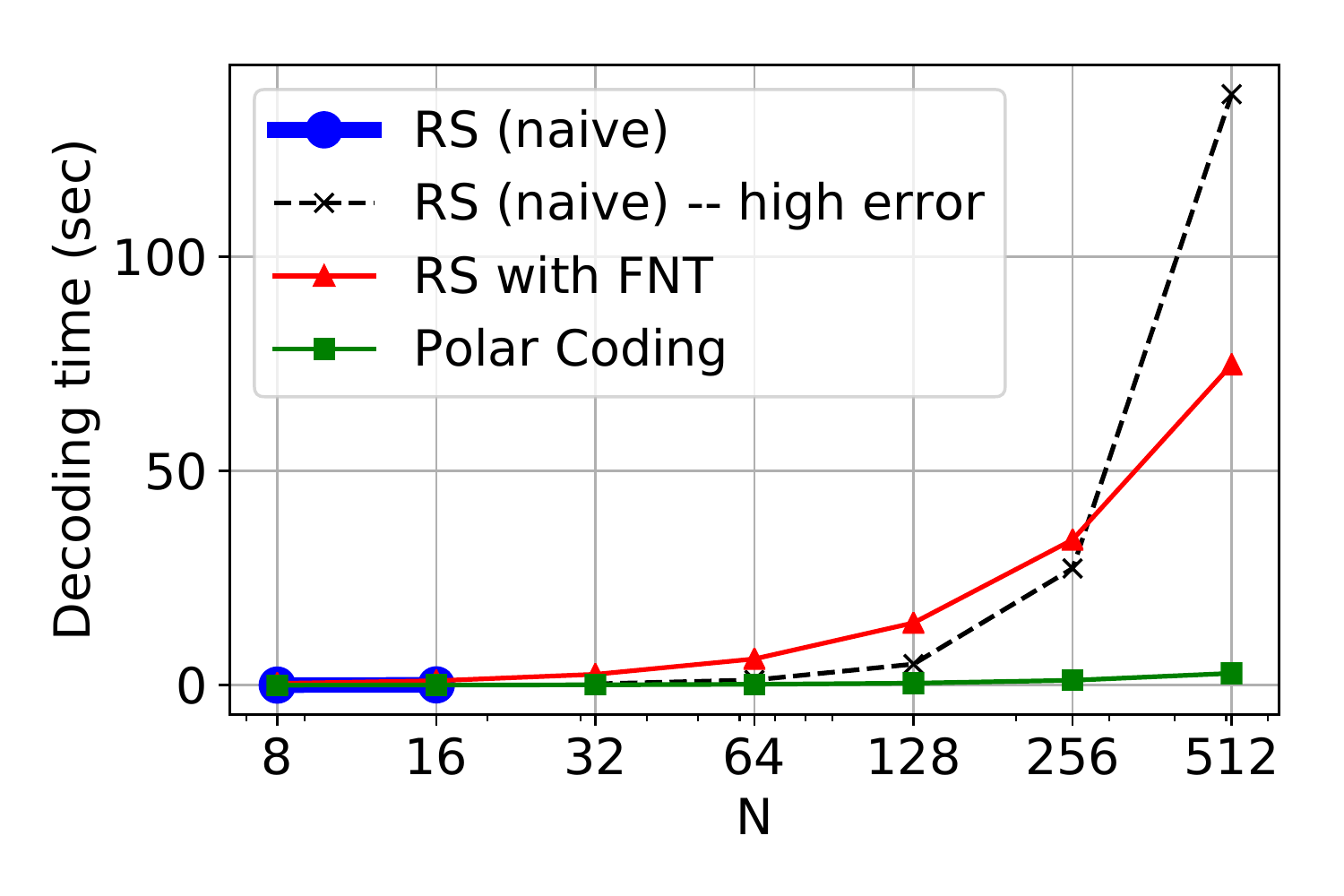}}
  \centerline{(b) Decoding}\medskip
\end{minipage}
\caption{Comparison of encoding and decoding speeds for RS and polar codes. Note that the horizontal axis is in log-scale.}
\label{coding_time_comparison_2}
\end{figure}

Figure \ref{coding_time_comparison_2} shows the time encoding and decoding algorithms take as a function of the number of workers $N$ for Reed-Solomon and polar codes. For Reed-Solomon codes, we implemented two encoders and decoders. The first approach is the naive approach where encoding is done using matrix multiplication ($O(N^2)$) and decoding is done by solving a linear system ($O(N^3)$), hence the naive encoder and decoder can support full-precision data. The second approach is the fast implementation for both encoding and decoding (of complexities $O(N\log N)$ and $O(N\log^2 N)$, respectively). The fast implementation is based on Fermat Number Transform (FNT), hence only supports finite field data. In obtaining the plots in Figure \ref{coding_time_comparison_2}, we always used $0.5$ as the rate and performed the computation $Ax$ where $A$ is $(100N \times 5000)$-dimensional and $x$ is $(5000\times 1000)$-dimensional. The line in Figure \ref{coding_time_comparison_2}(b) with markers 'x' and dashed lines indicates that the error due to the decoder is unacceptably high (this can happen because we used full-precision data for the naive RS decoder).

Figure \ref{coding_time_comparison_2} illustrates that polar codes take much less time for encoding and especially decoding compared to Reed-Solomon codes. This is because of the constants hidden in the complexities of fast decoders for Reed-Solomon decoders which is not the case for polar codes. We note that it might be more advantageous to use RS codes for small $N$ values because they have the MDS properties and can encode and decode fast enough for small $N$. However, in serverless computing where each function has limited resources and hence using large $N$ values is usually the case, we need faster encoding and decoding algorithms. Considering the comparison in Figure \ref{coding_time_comparison_2}, polar codes are more suitable to use with serverless computing.


\subsection{Polarized Computation Times}
\begin{figure}[htb]
\begin{minipage}[b]{0.48\linewidth}
  \centering
  \centerline{\includegraphics[width=4.5cm]{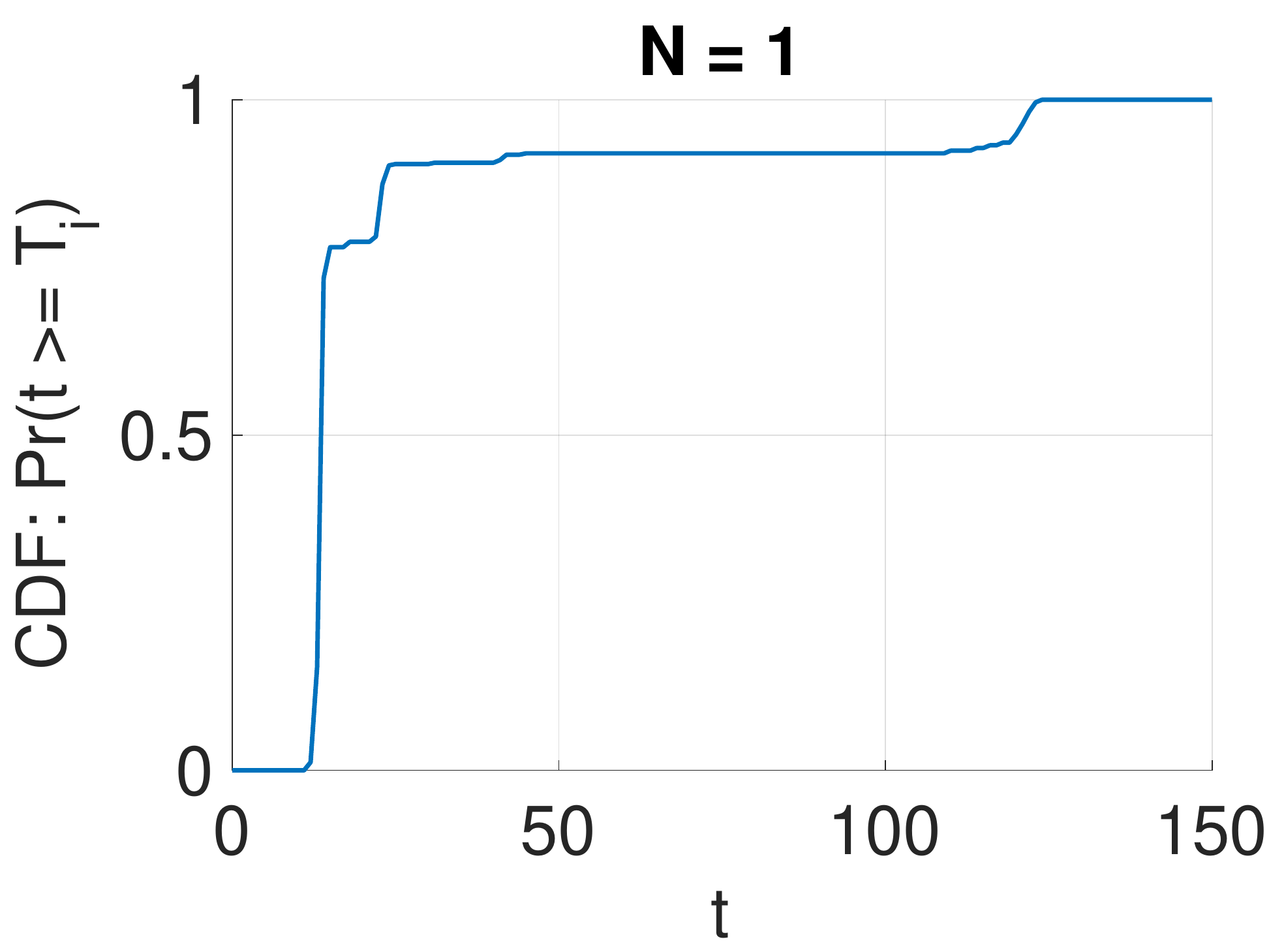}}
  \centerline{(a) $N=1$}\medskip
\end{minipage}
\hfill
\begin{minipage}[b]{0.48\linewidth}
  \centering
  \centerline{\includegraphics[width=4.5cm]{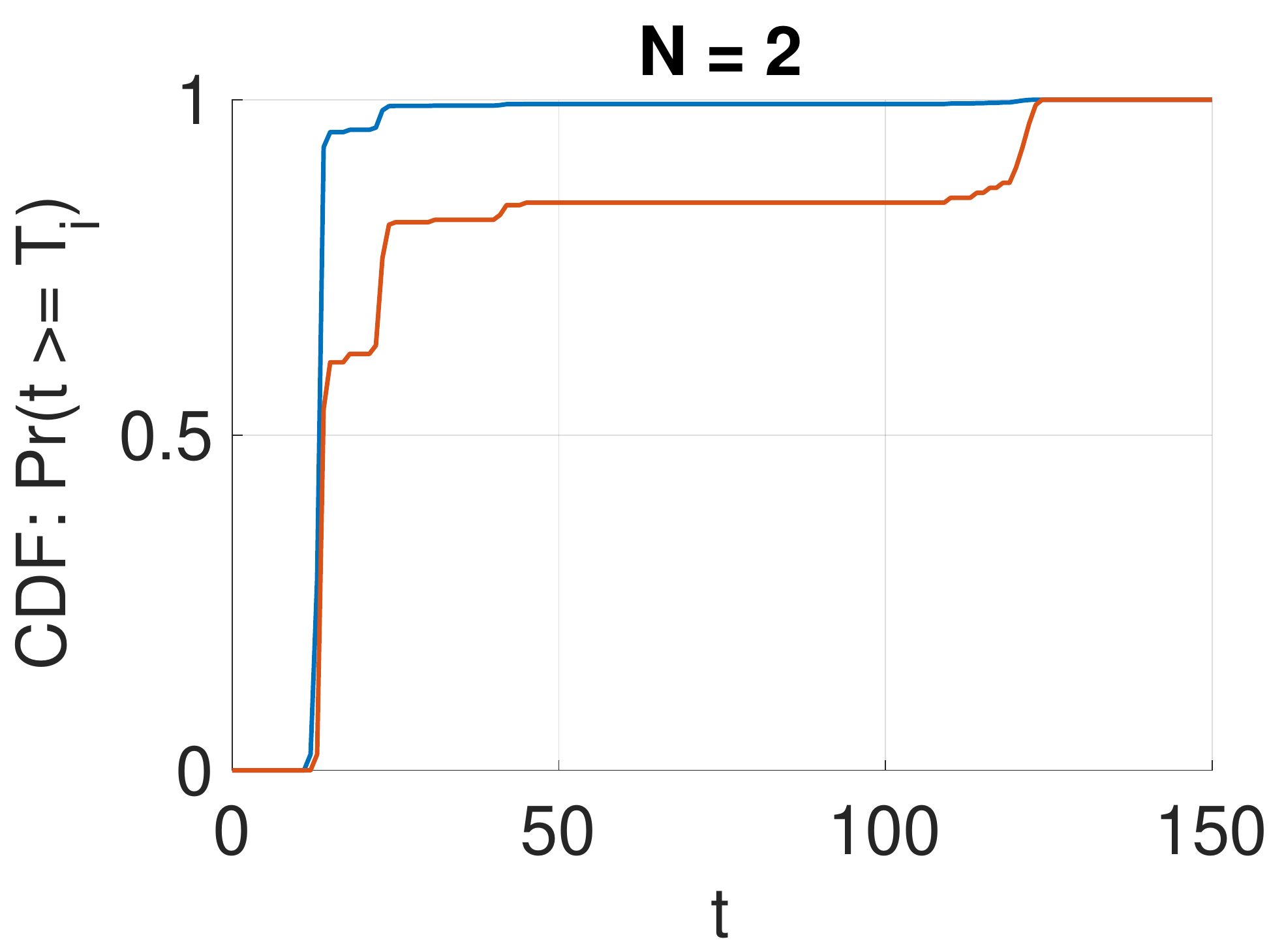}}
  \centerline{(b) $N=2$}\medskip
\end{minipage}
\hfill
\begin{minipage}[b]{.48\linewidth}
  \centering
  \centerline{\includegraphics[width=4.5cm]{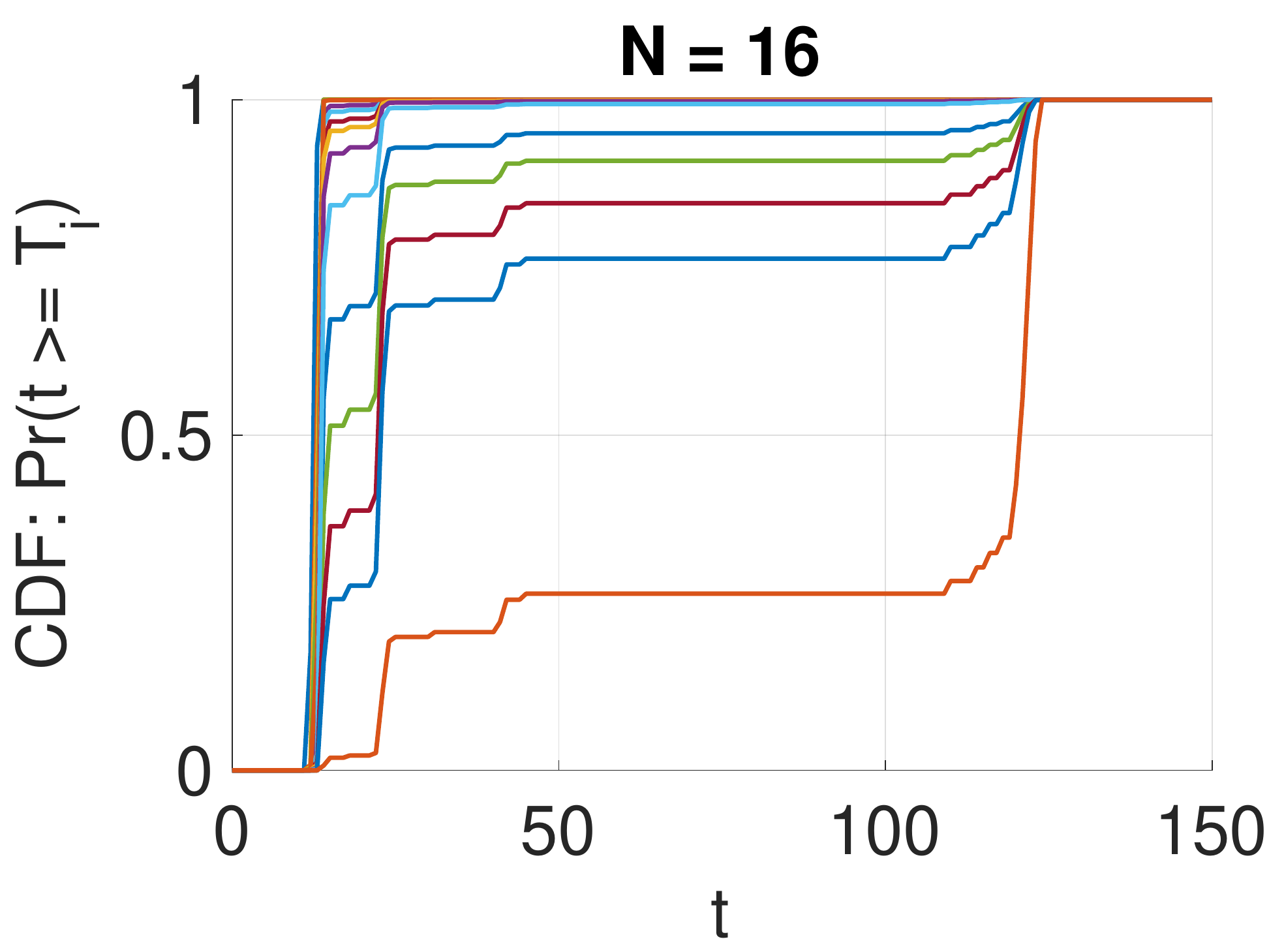}}
  \centerline{(c) $N=16$}\medskip
\end{minipage}
\hfill
\begin{minipage}[b]{0.48\linewidth}
  \centering
  \centerline{\includegraphics[width=4.5cm]{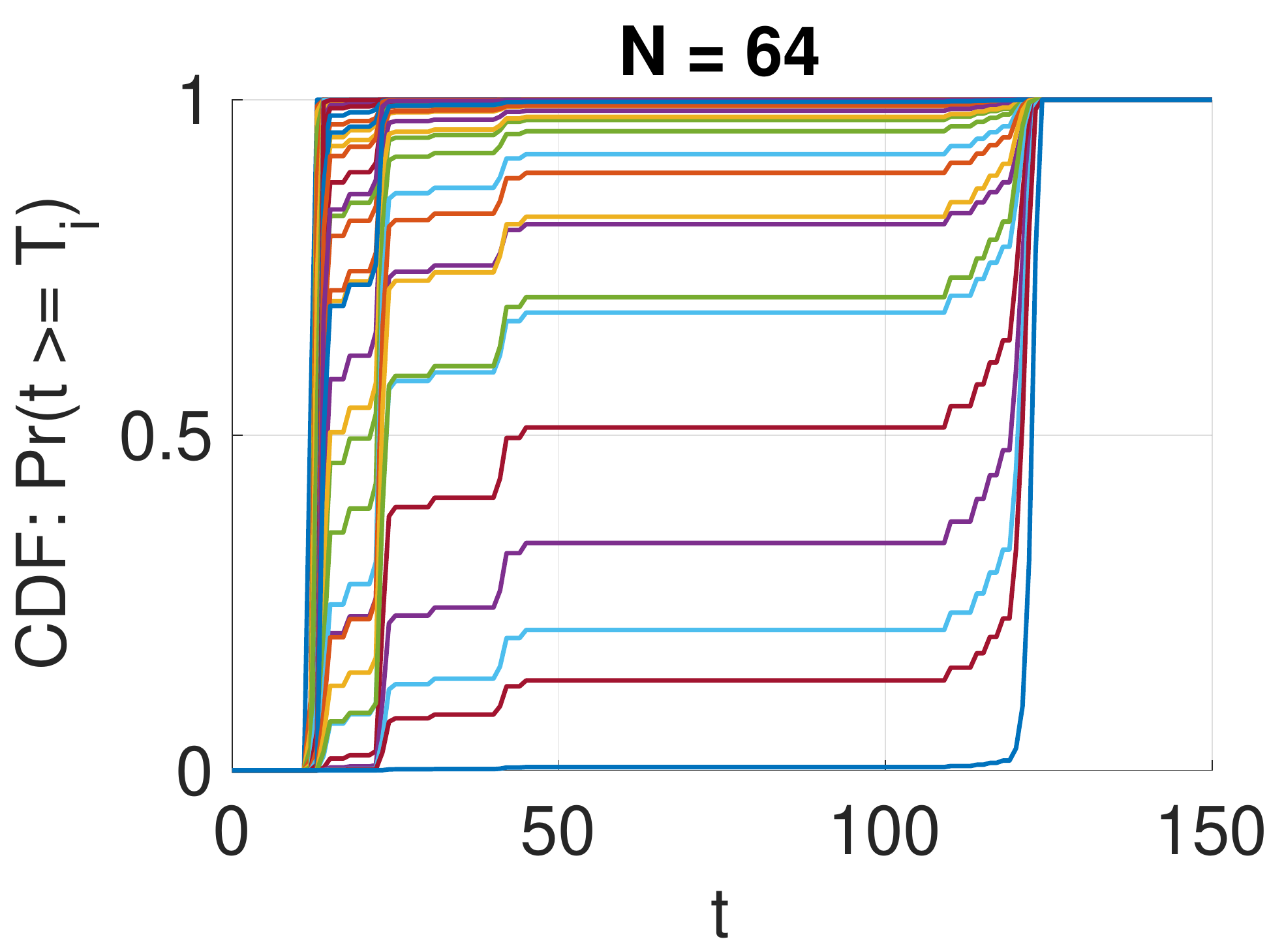}}
  \centerline{(d) $N=64$}\medskip
\end{minipage}
\caption{Polarization of empirical CDFs.}
\label{polarized_cdfs}
\end{figure}

Figure \ref{polarized_cdfs} illustrates how the empirical CDFs of computation times polarize as we increase the number of functions $N$. Figure \ref{polarized_cdfs}(a) shows the empirical CDF of the computation times of AWS Lambda functions, obtained by timing $500$ AWS Lambda functions running the same Python program. The other plots in Figure \ref{polarized_cdfs} are generated assuming that there are $N$ functions with i.i.d. run time distributions (the CDF of which is plotted in Figure \ref{polarized_cdfs} (a)), and show the resulting transformed CDFs of the polarized computation times. Note that the empirical CDFs move away from each other (i.e. polarize) as $N$ is increased and we obtain \textit{better} and \textit{worse} run time distributions. Note that the freezing operation in the code construction corresponds to not using workers with worse run times.

\subsection{Empirical Distribution of Decodability Time}
\begin{figure}[ht!]
\begin{minipage}[b]{0.3\linewidth}
  \centering
  \centerline{\includegraphics[width=3cm]{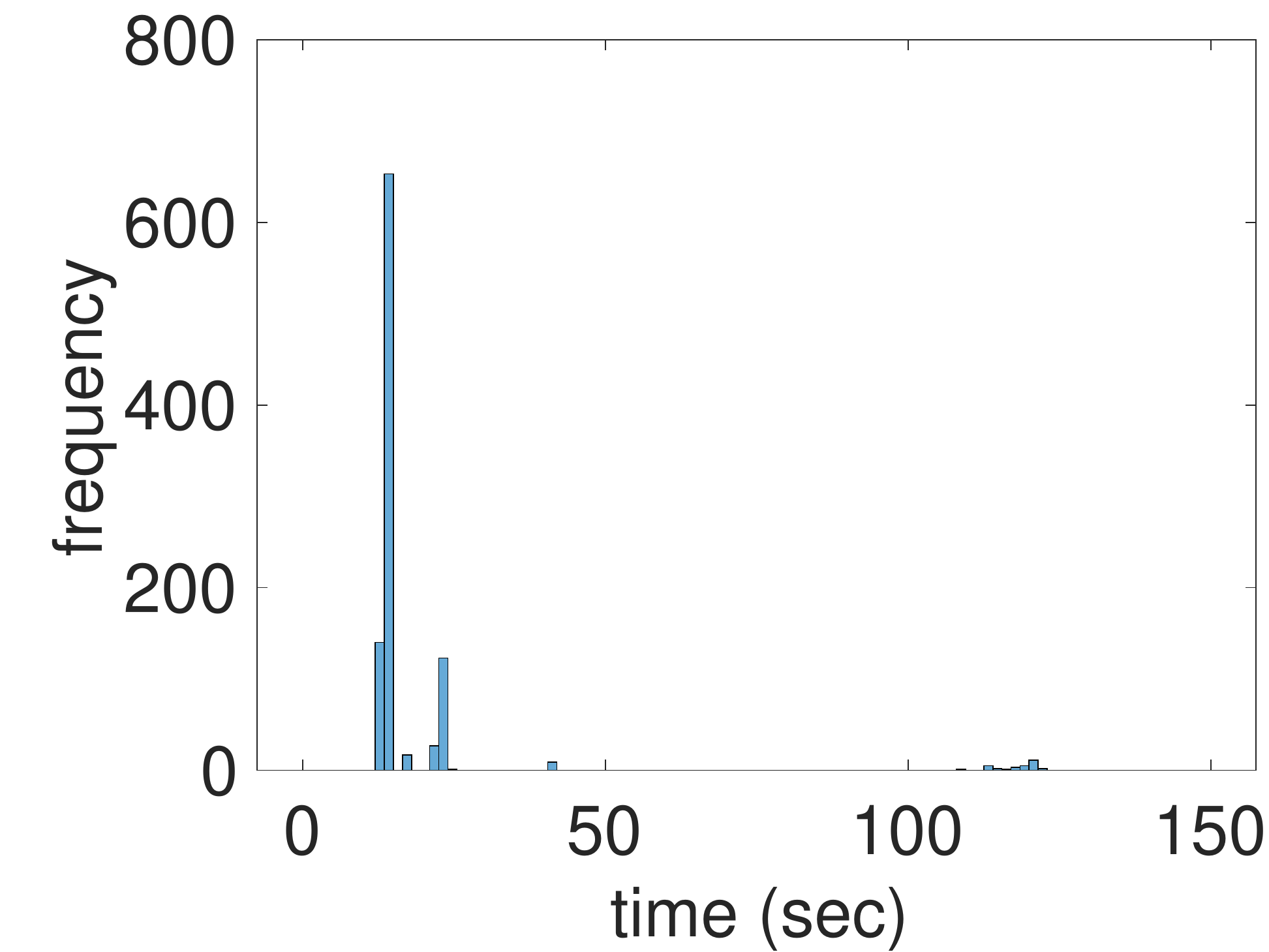}}
  \centerline{(a) Polar, $N=8$}\medskip
\end{minipage}
\hfill
\begin{minipage}[b]{0.3\linewidth}
  \centering
  \centerline{\includegraphics[width=3cm]{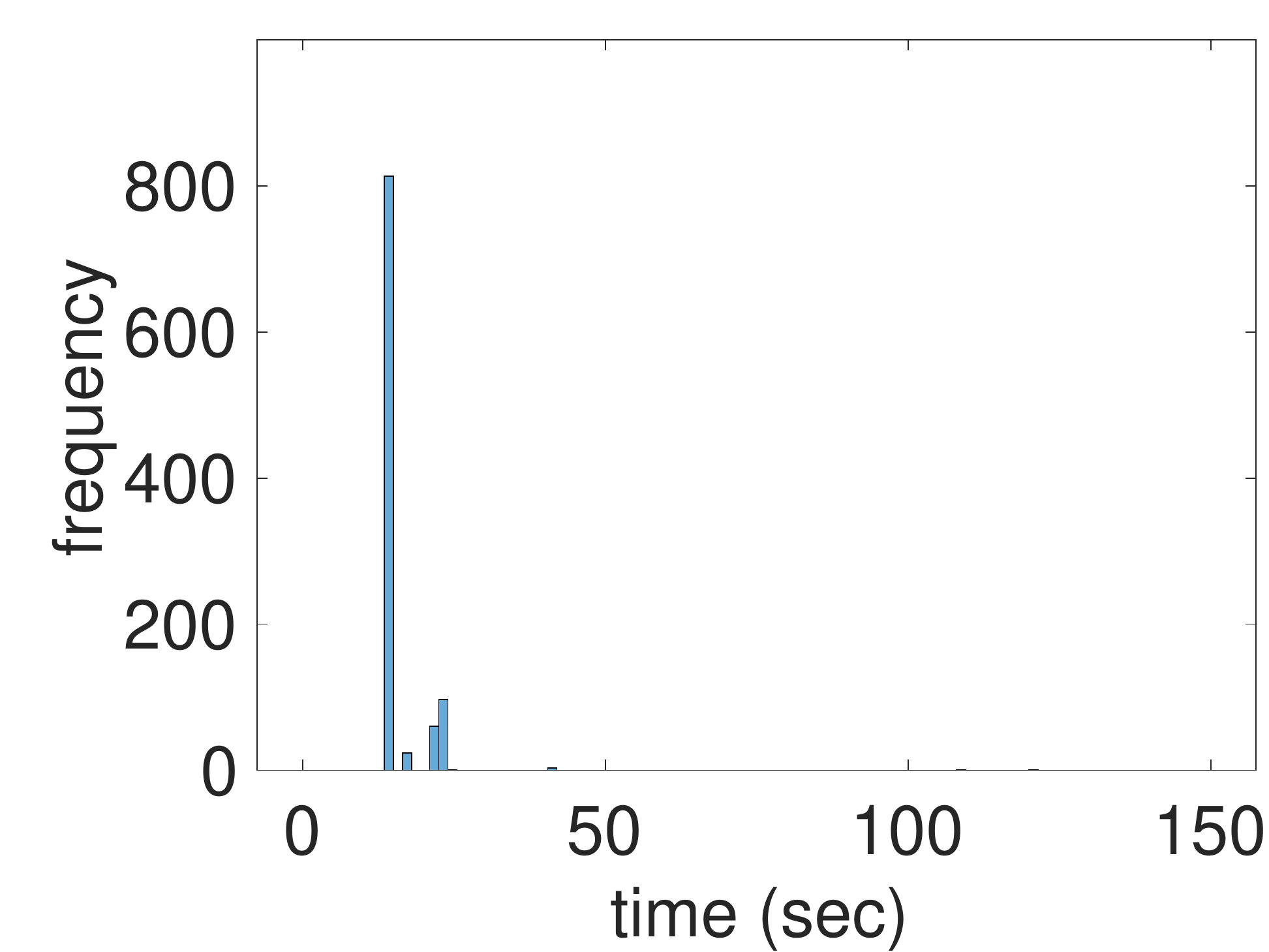}}
  \centerline{(b) Polar, $N=64$}\medskip
\end{minipage}
\hfill
\begin{minipage}[b]{0.3\linewidth}
  \centering
  \centerline{\includegraphics[width=3cm]{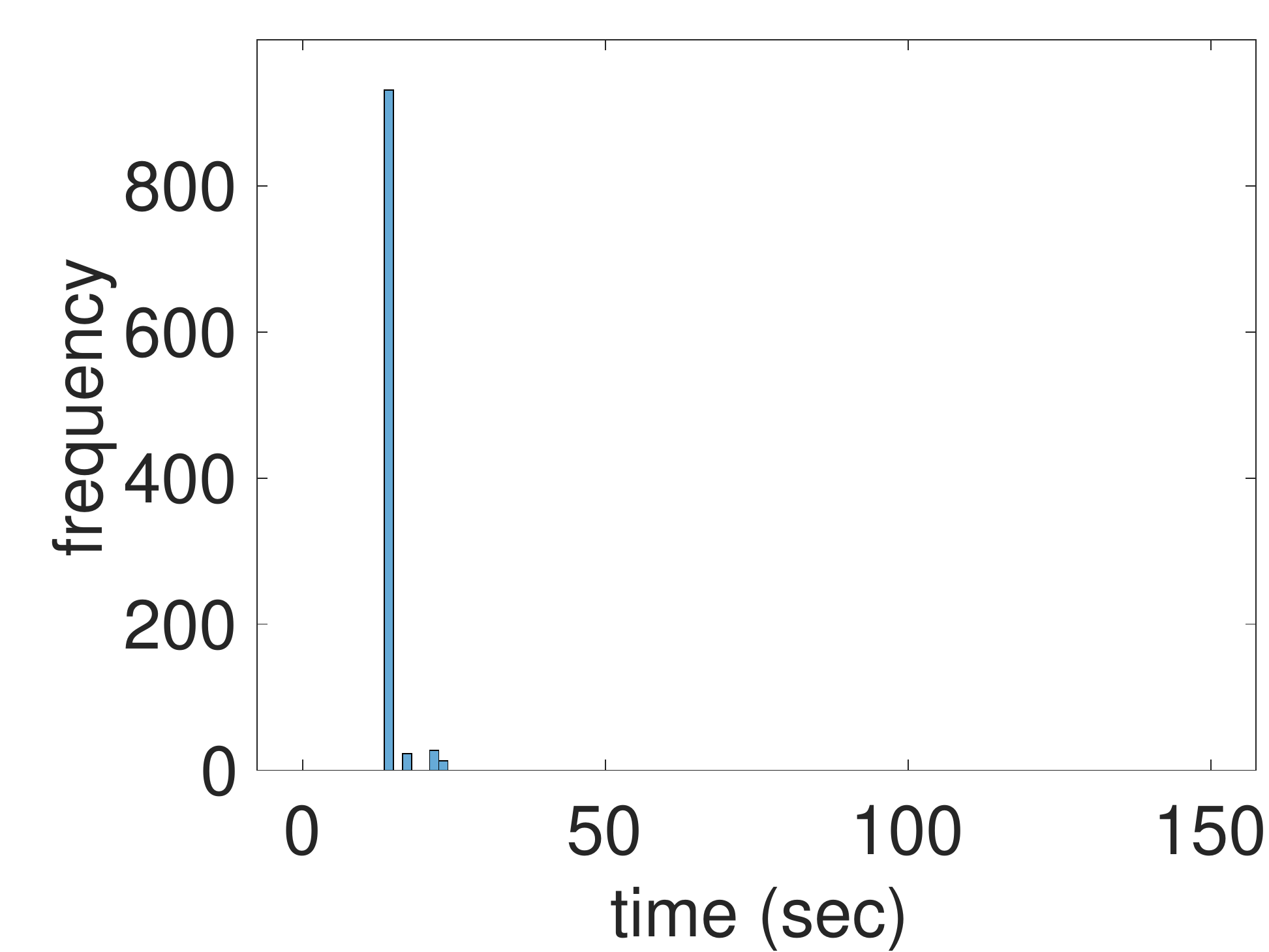}}
  \centerline{(c) Polar, $N=512$}\medskip
\end{minipage}
\hfill
\begin{minipage}[b]{0.3\linewidth}
  \centering
  \centerline{\includegraphics[width=3cm]{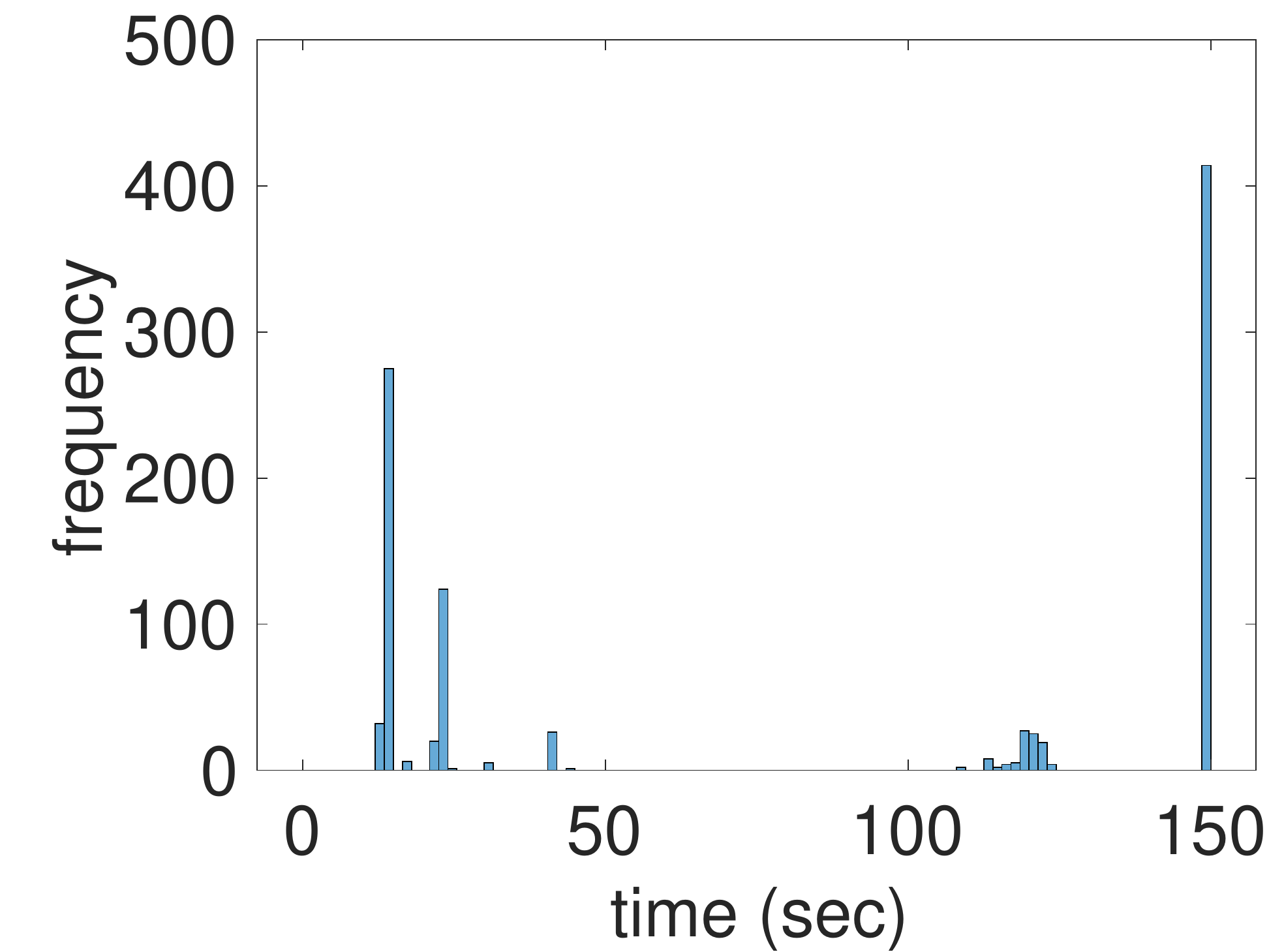}}
  \centerline{(d) LT, $N=8$}\medskip
\end{minipage}
\hfill
\begin{minipage}[b]{0.3\linewidth}
  \centering
  \centerline{\includegraphics[width=3cm]{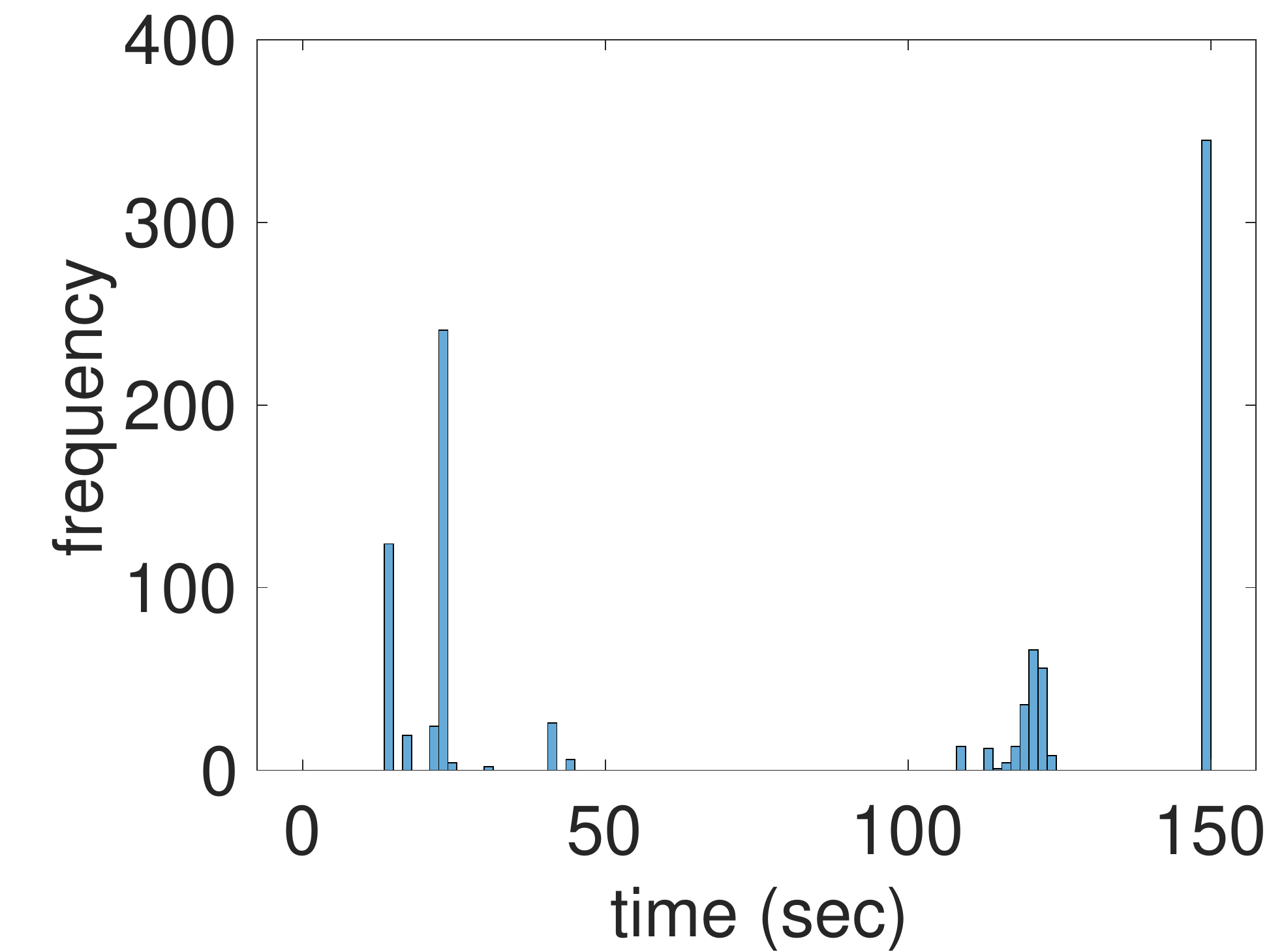}}
  \centerline{(e) LT, $N=64$}\medskip
\end{minipage}
\hfill
\begin{minipage}[b]{0.3\linewidth}
  \centering
  \centerline{\includegraphics[width=3cm]{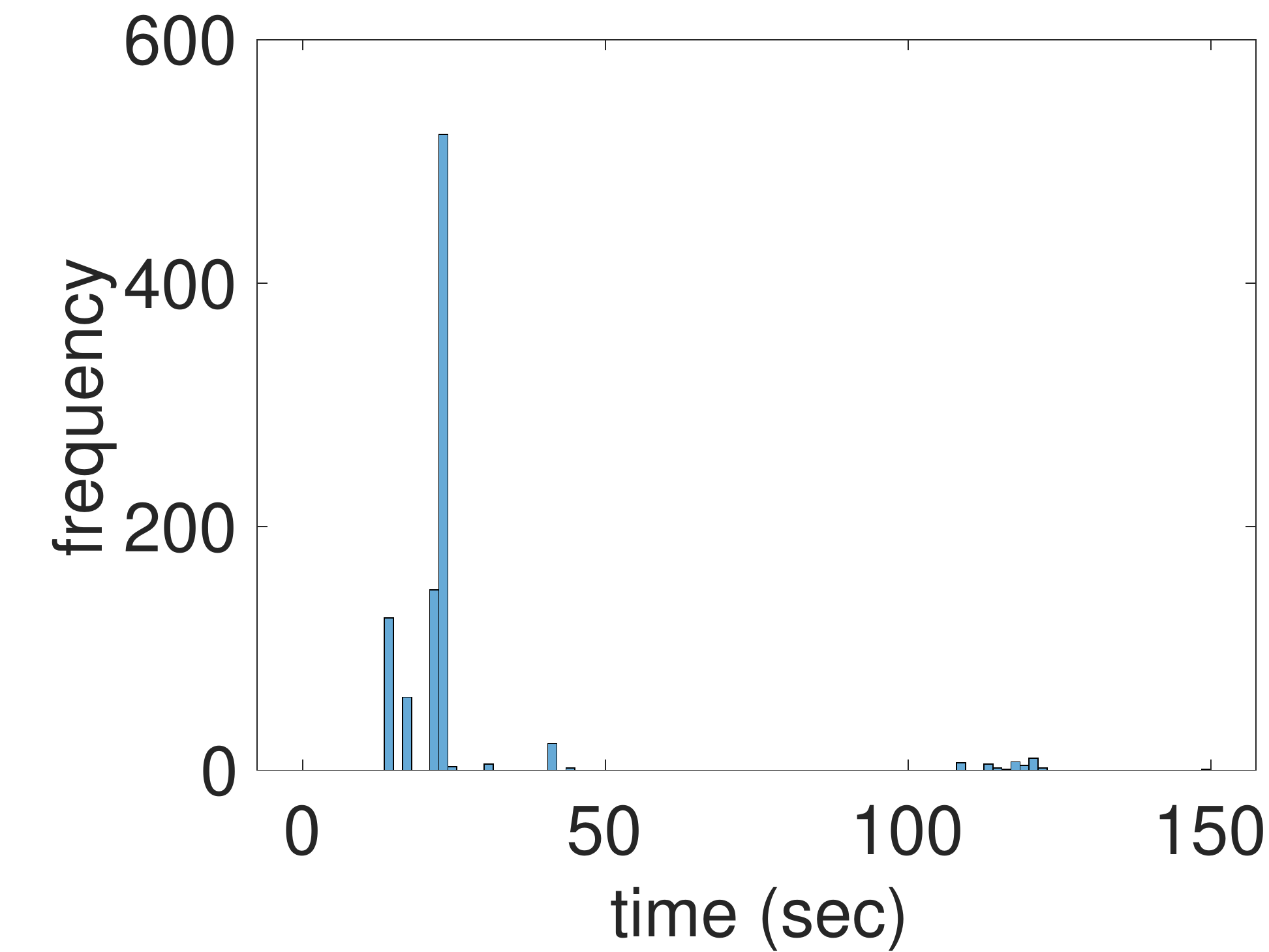}}
  \centerline{(f) LT, $N=512$}\medskip
\end{minipage}
\hfill
\begin{minipage}[b]{0.3\linewidth}
  \centering
  \centerline{\includegraphics[width=3cm]{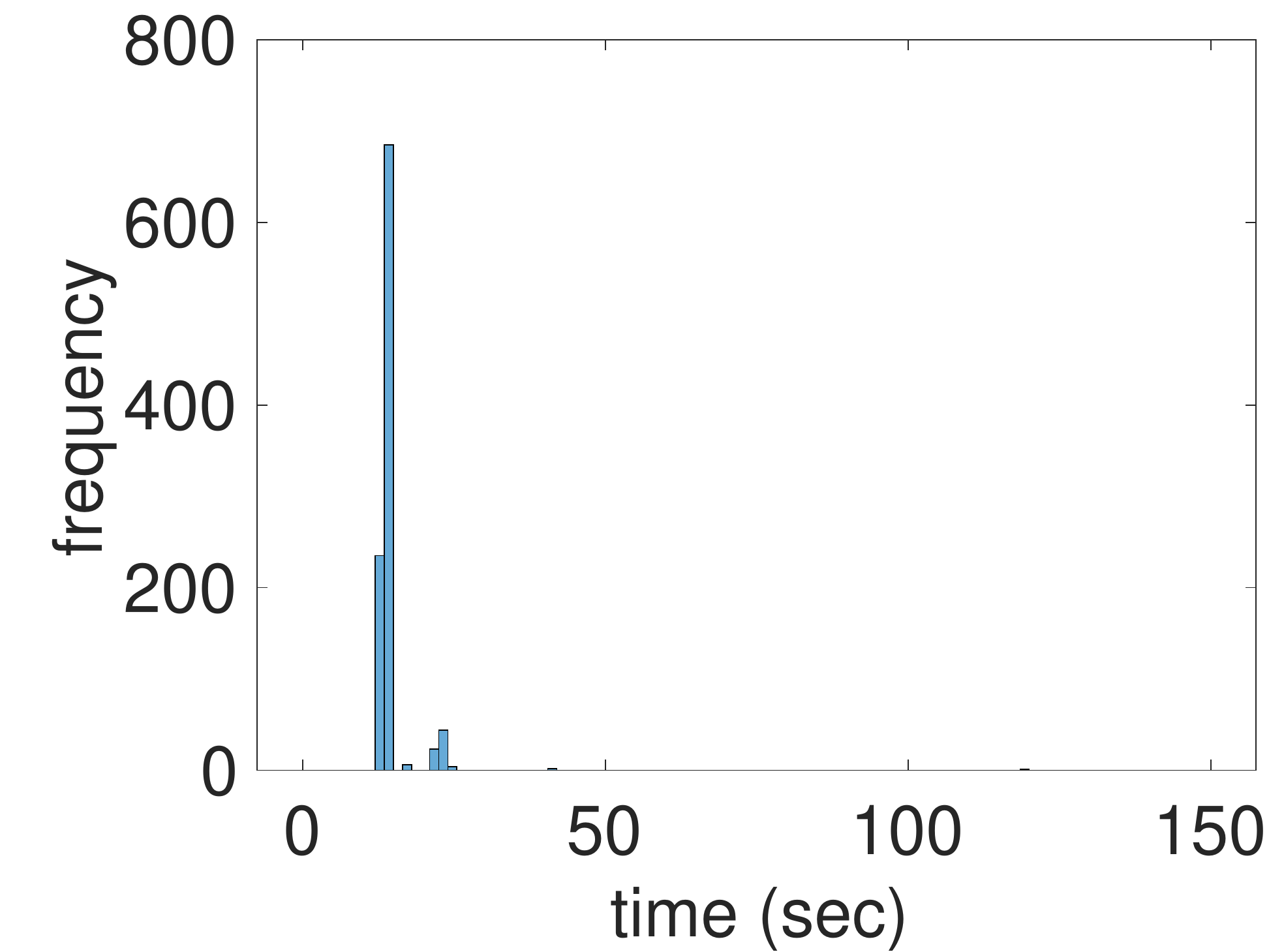}}
  \centerline{(g) MDS, $N=8$}\medskip
\end{minipage}
\hfill
\begin{minipage}[b]{0.3\linewidth}
  \centering
  \centerline{\includegraphics[width=3cm]{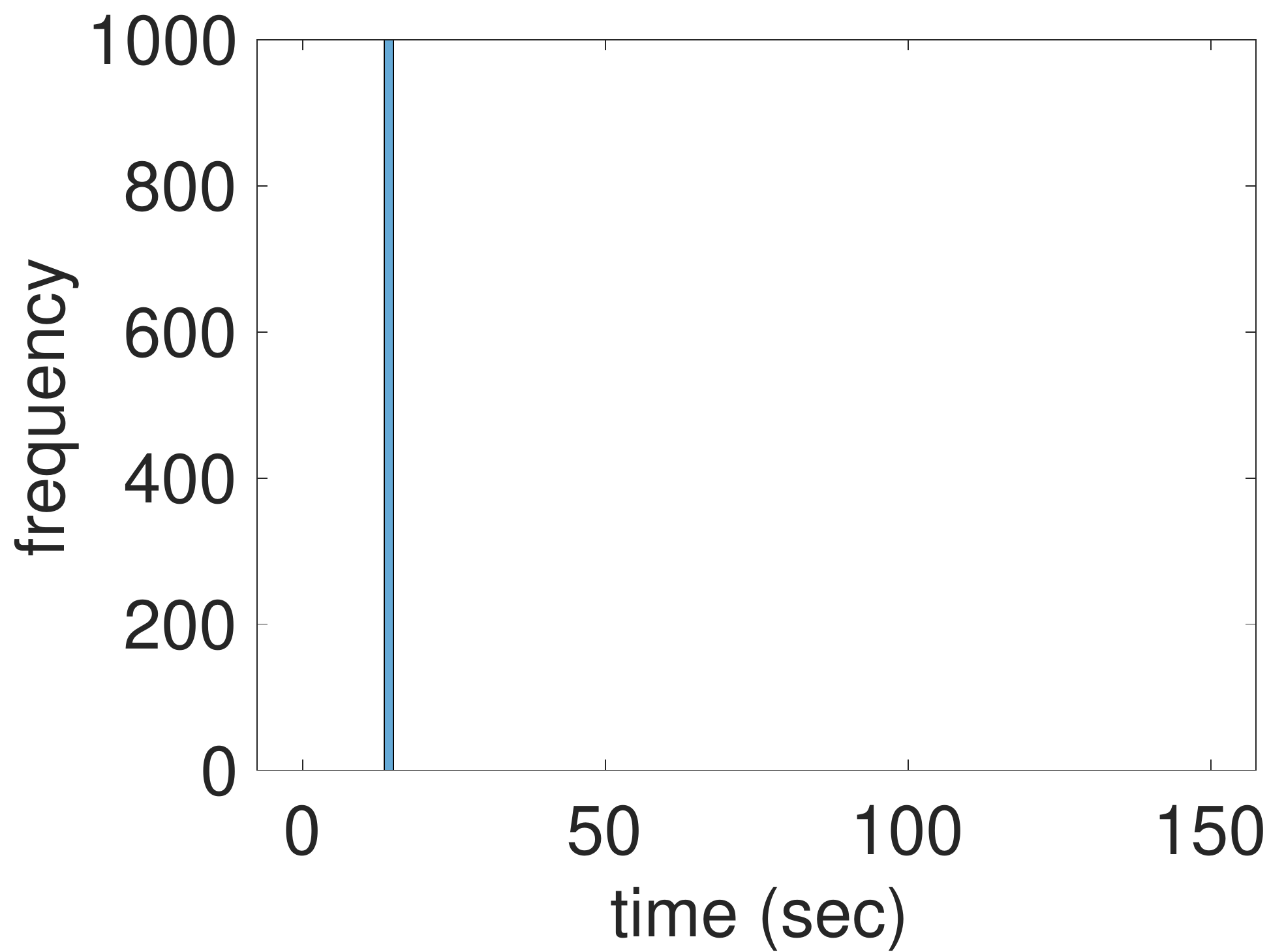}}
  \centerline{(h) MDS, $N=64$}\medskip
\end{minipage}
\hfill
\begin{minipage}[b]{0.3\linewidth}
  \centering
  \centerline{\includegraphics[width=3cm]{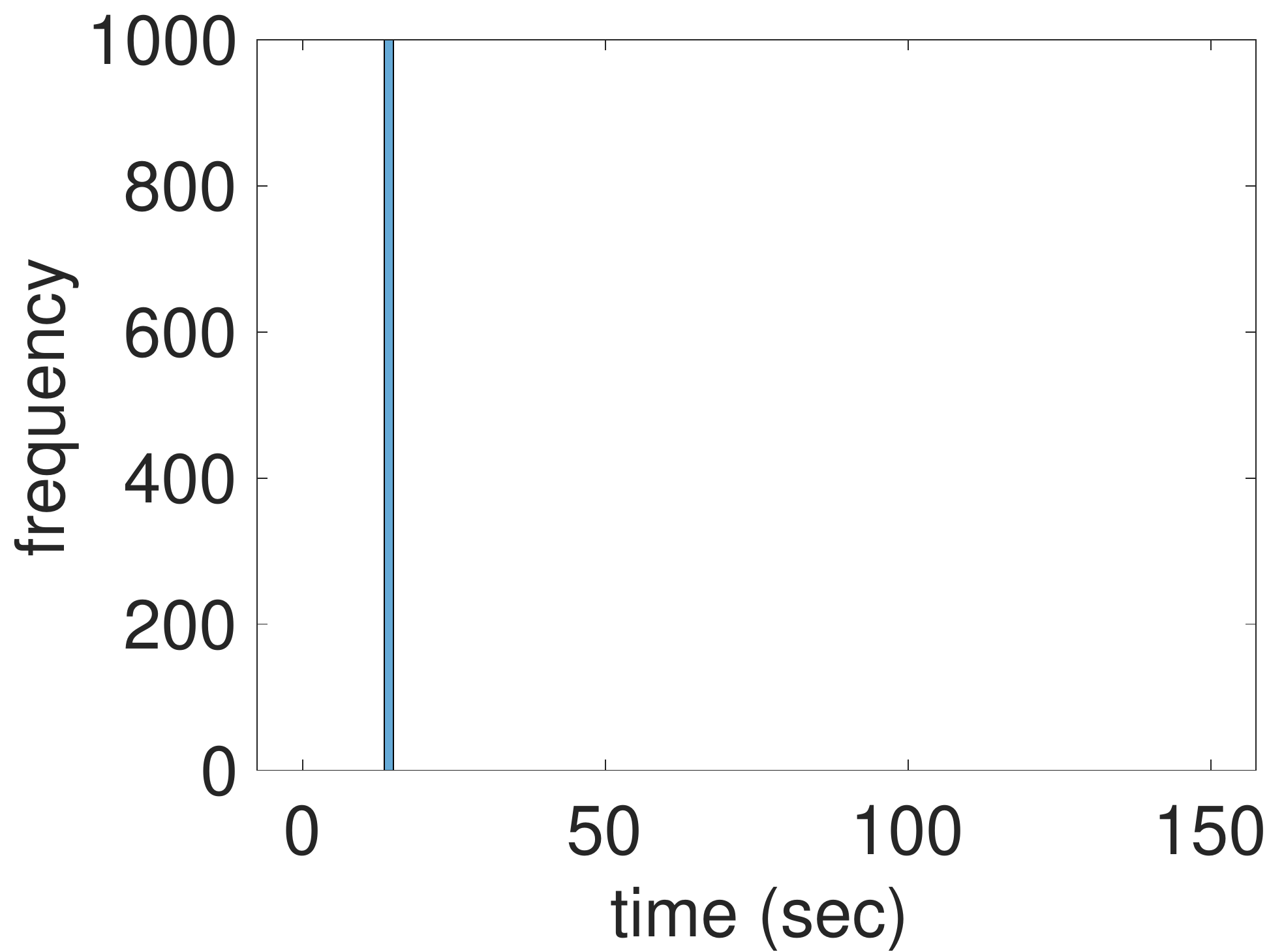}}
  \centerline{(i) MDS, $N=512$}\medskip
\end{minipage}
\caption{Histograms of decodability time for polar, LT, and MDS codes.}
\label{decod_time_histograms}
\end{figure}

We refer to the time instance where the available outputs become decodable for the first time as \textit{decodability time}. Figure \ref{decod_time_histograms} shows the histograms of the decodability time for different values of $N$ for polar, LT, and MDS codes, respectively. These histograms were obtained by sampling i.i.d. worker run times with replacement from the input distribution whose CDF is plotted in Figure \ref{polarized_cdfs}(a) and by repeating this $1000$ times. Further, $\epsilon=0.375$ was used as the erasure probability. We observe that as $N$ increases, the distributions converge to the dirac delta function, showing that for large $N$ values, the decodability time becomes deterministic.

Plots in Figure \ref{decod_time_histograms}(d,e,f) are the decodability time histograms for LT codes with peeling decoder. The degree distribution is the robust soliton distribution as suggested in \cite{mallick2018rateless}. We see that polar codes achieve better decodability times than LT codes. Plots in Figure \ref{decod_time_histograms}(g,h,i) on the other hand show that MDS codes perform better than polar codes in terms of decodability time, which is expected. When considering this result, one should keep in mind that for large $N$, MDS codes take much longer times to encode and decode compared to polar codes as we discussed previously. In addition, we see that for large $N$ values, the gap between the decodability time performances closes.


\begin{figure}[htb]
\begin{minipage}[b]{0.48\linewidth}
  \centering
  \centerline{\includegraphics[width=3cm]{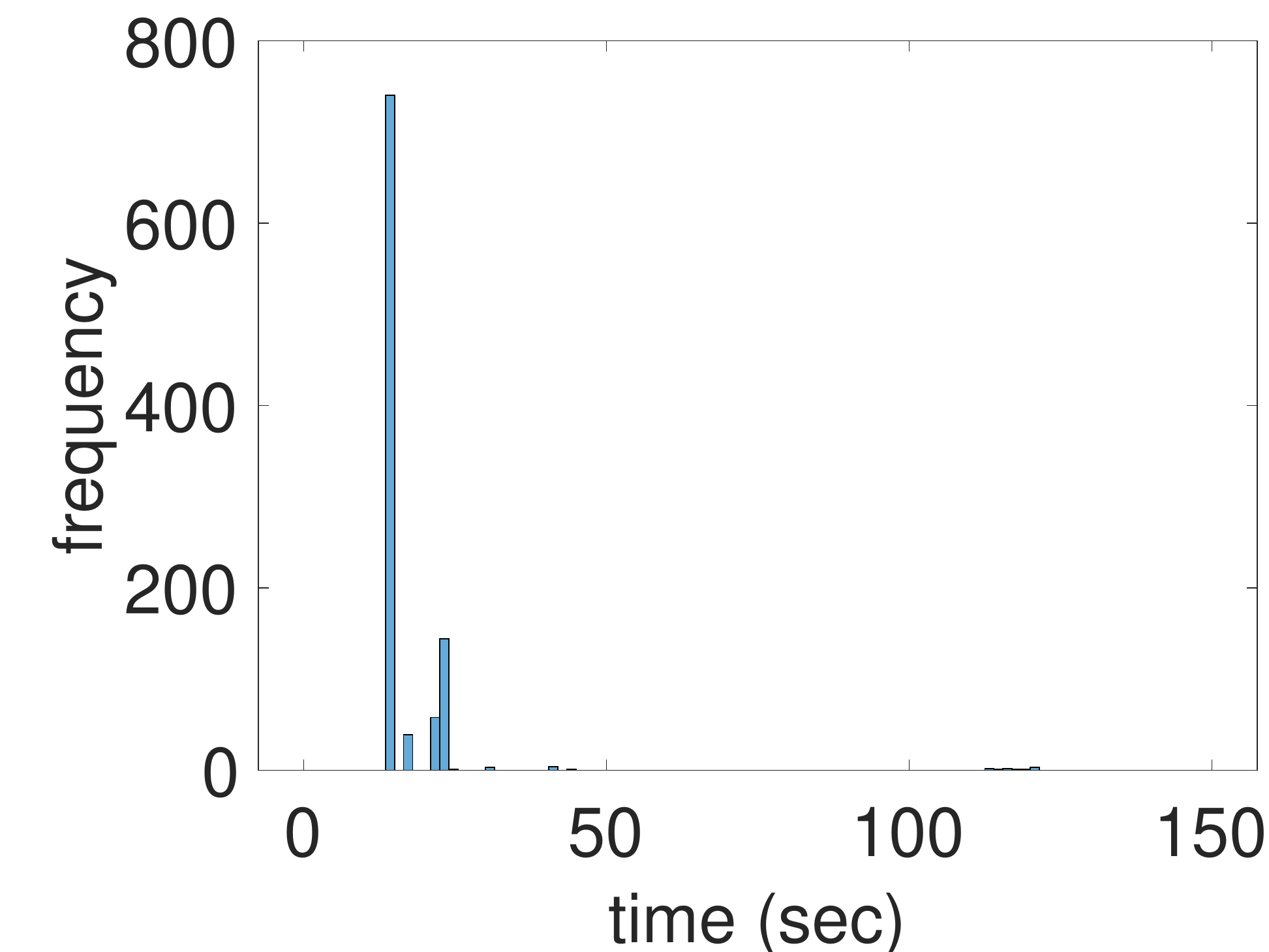}}
  \centerline{(a) Polar, $N=64$}\medskip
\end{minipage}
\hfill
\begin{minipage}[b]{0.48\linewidth}
  \centering
  \centerline{\includegraphics[width=3cm]{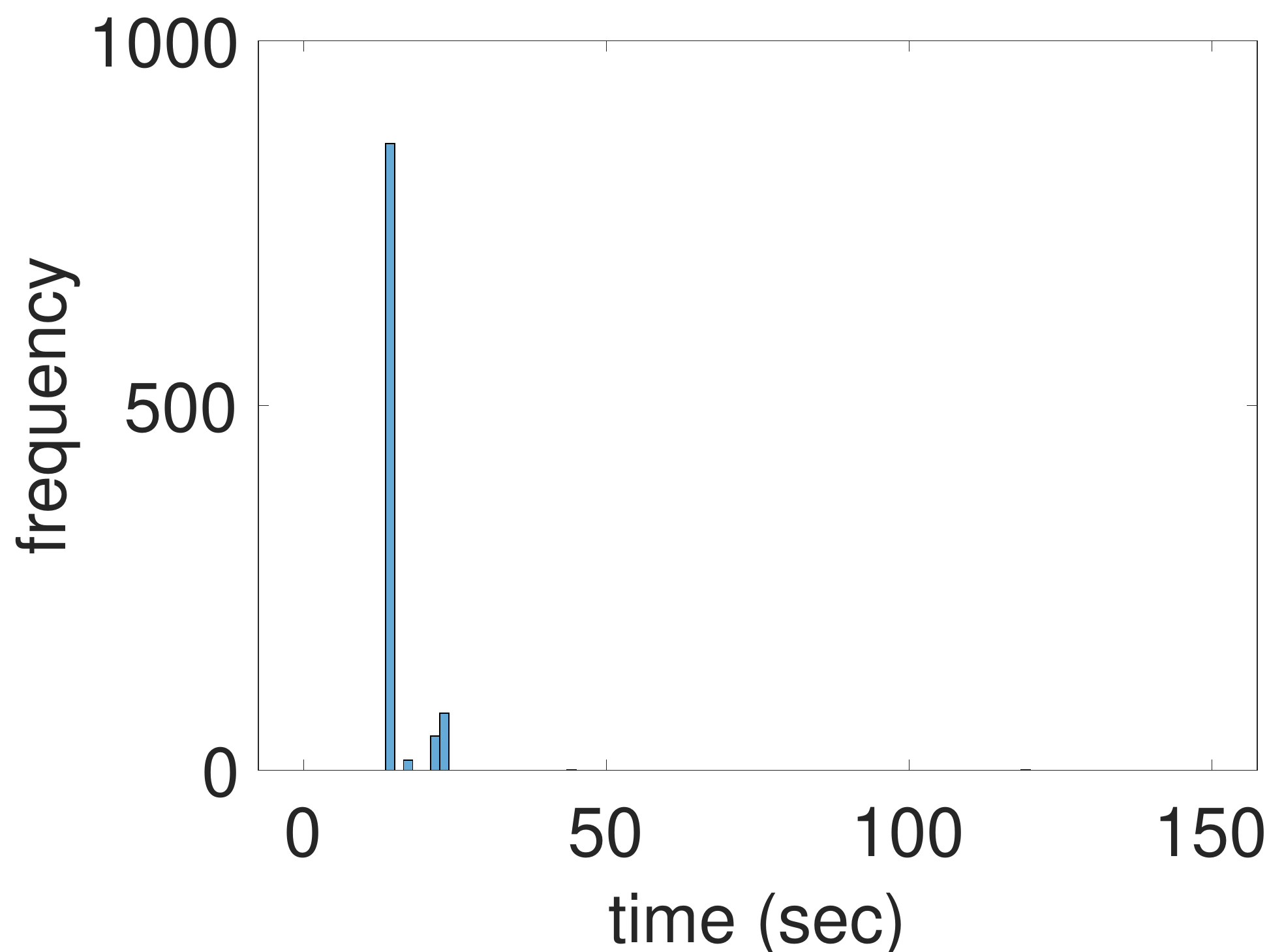}}
  \centerline{(b) Polar, $N=256$}\medskip
\end{minipage}
\hfill
\begin{minipage}[b]{.48\linewidth}
  \centering
  \centerline{\includegraphics[width=3cm]{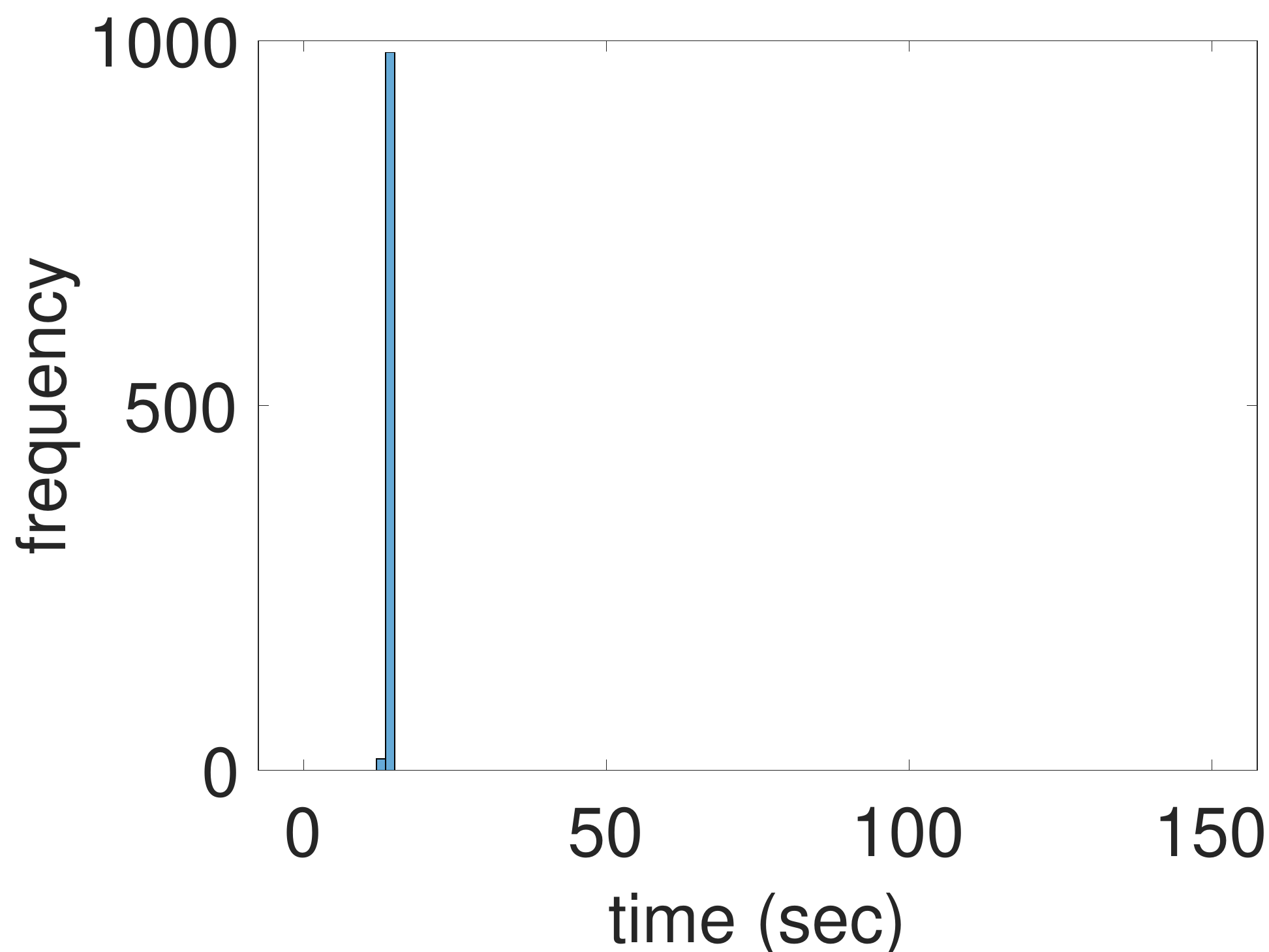}}
  \centerline{(c) MDS, $N=64$}\medskip
\end{minipage}
\hfill
\begin{minipage}[b]{0.48\linewidth}
  \centering
  \centerline{\includegraphics[width=3cm]{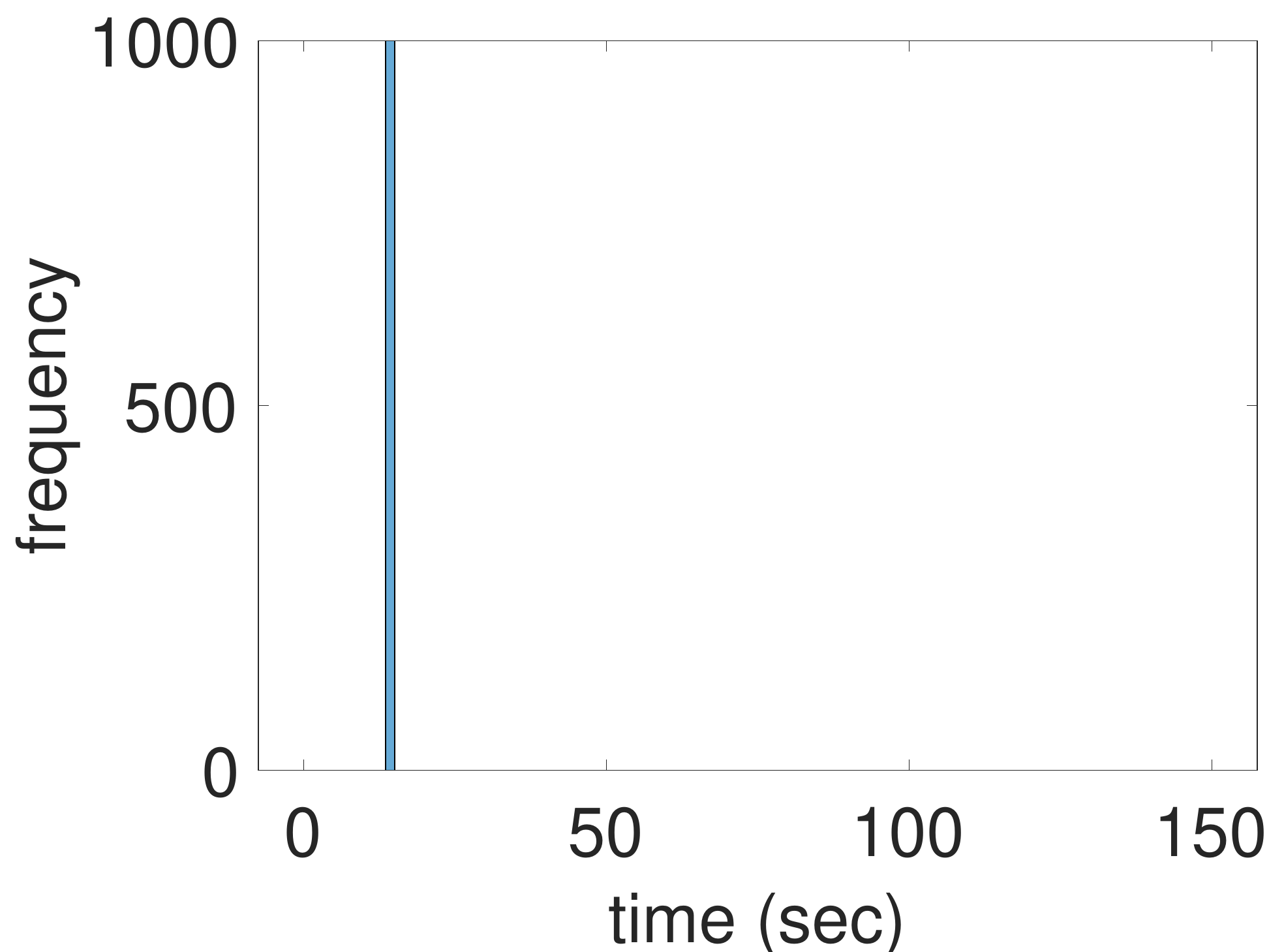}}
  \centerline{(d) MDS, $N=256$}\medskip
\end{minipage}
\caption{Histograms of decodability time for polar codes and MDS codes for matrix multiplication.}
\label{decod_time_histograms_2D}
\end{figure}

Figure \ref{decod_time_histograms_2D} shows decodability time histograms for computing the matrix multiplication $AB$ for polar (using the 2D decoder outlined in Alg. \ref{2d_decoder}) and MDS codes. The rate for encoding $A$ is $\frac{3}{4}$ and it is the same for $B$, which gives an overall rate of $\frac{9}{16}$. The decodability time for polar codes is the first time instance in which the decoder in Alg. \ref{2d_decoder} is able to decode the available set of outputs. The decodability time for MDS codes is whenever $\frac{9N}{16}$ of the outputs is available. Figure \ref{decod_time_histograms_2D} illustrates that the decodability time for the polar codes is only slightly worse than for MDS codes. Note that the decodability time for MDS codes is optimal since the recovery threshold is optimal for MDS codes when $A$ is partitioned row-wise and $B$ is partitioned column-wise when computing $AB$ \cite{yu2017polycode}. For $N=256$, we see in Figure \ref{decod_time_histograms_2D} that the gap between the decodability time performances of polar and MDS codes is negligibly small. Hence, as the number of functions is increased, the decodability time becomes no longer an important disadvantage of using polar codes instead of MDS codes.

\subsection{Coded Computation on AWS Lambda}
We have implemented the proposed coded computation scheme for the AWS Lambda platform and now present the results illustrating its performance. The implementation is written in Python. The compute nodes are AWS Lambda functions each with a memory of $1536$ MB and a timeout limit of $300$ seconds. We have utilized the Pywren module \cite{jonas2017pywren} for running code in AWS Lambda. The master node in our implementation is a Macbook Pro laptop with $8$ GB of RAM.

\begin{figure}[htbp]
  \centering
  \includegraphics[width=0.48\textwidth]{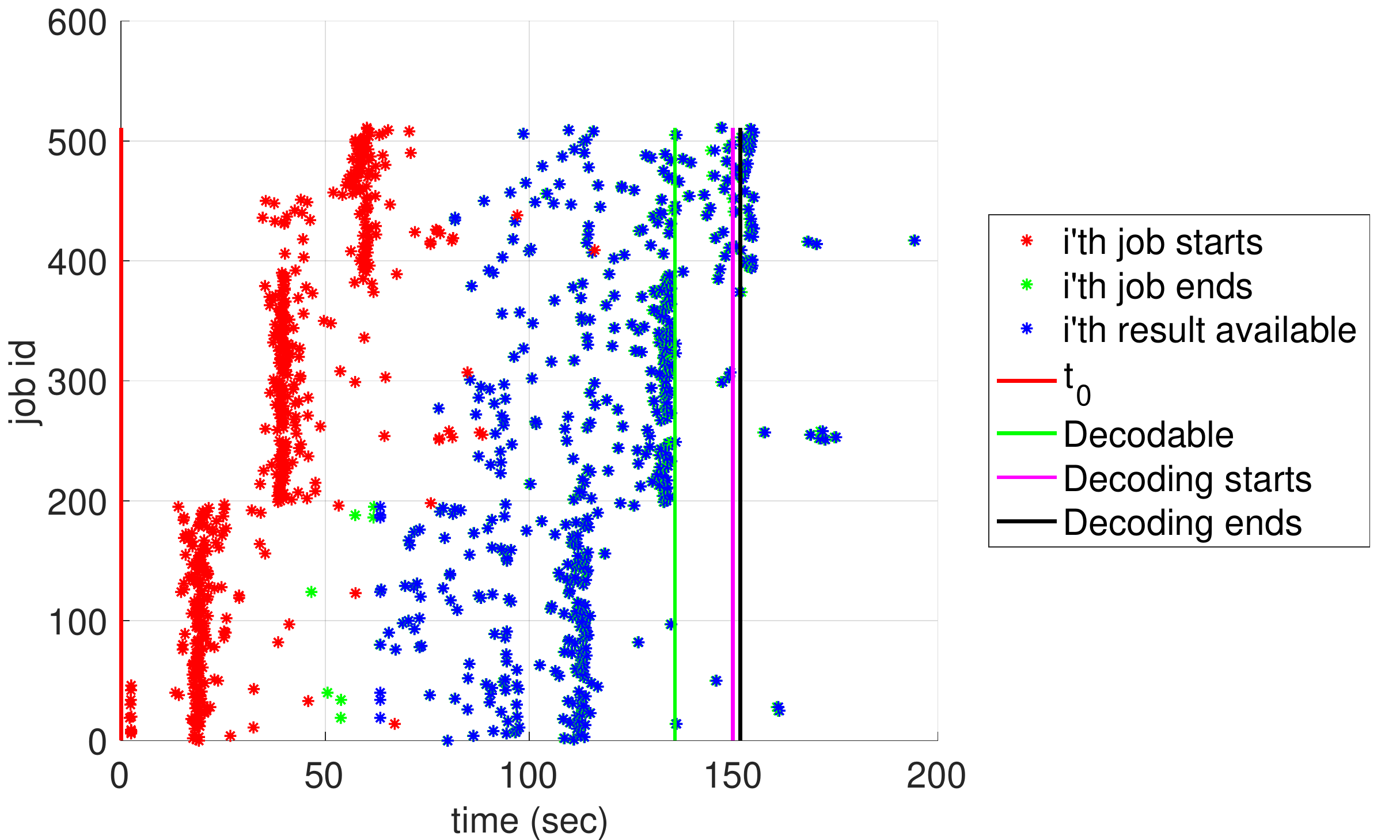}
  \caption{Job output times and decoding times for $N=512$ using AWS Lambda.}
  \label{plot_times_N_512}
\end{figure}

Given in Figure \ref{plot_times_N_512} is a plot showing the timings of serverless functions, and of the decoding process, where we compute $A\times x$ where $A \in \mathbb{R}^{38400\times 3000000}$ and $x\in \mathbb{R}^{3000000\times 20}$ using $N=512$ serverless functions, and taking $\epsilon = 0.25$. The vertical axis represents the numbers we assign to the serverless functions (i.e. job id). Encoding process is not included in the plot, as it is usually the case for many applications that we multiply the same $A$ matrix with different $x$'s. Hence for an application that requires iterative matrix multiplications with the same $A$ but different $x$'s, it is sufficient to encode $A$ once, but decoding has to be repeated for every different $x$. Therefore, the overall time would be dominated by the matrix multiplications and decoding, which are repeated many times unlike encoding which takes place once.

For the experiment shown in Figure \ref{plot_times_N_512}, each serverless function instance performs a matrix multiplication with dimensions $(100\times 3000000)$ and $(3000000\times 20)$. A matrix with dimensions $(100\times 3000000)$ requires a memory of $2.3$ GB. It is not feasible to load a matrix of this size into the memory at once; hence, each serverless function loads $1/10$'th of its corresponding matrix at a time. After computing $1/10$'th of its overall computation, it moves on to perform the next $1/10$ of its computation. In Figure \ref{plot_times_N_512}, the red stars show the times each functions starts running. Green stars (mostly invisible because they are covered by the blue stars) show the times functions finish running. Blue stars indicate the times that the outputs of functions become available. Green (vertical) line is the first time the available outputs become decodable. After the available outputs become decodable, we cannot start decoding right away; we have to wait for their download (to the master node) to be finished, and this is shown by the magenta line. Magenta line also indicates the beginning of the decoding process, and lastly the black line indicates that decoding is finished.

We observe that there are straggling AWS Lambda functions that finish after decoding is over, showing the advantage of our coded computation scheme over an uncoded scheme where we would have to wait for all AWS Lambda functions to finish their assigned tasks. The slowest serverless function seems to have finished around $t=200$, whereas the result is available at around $t=150$ for the proposed coded computation. 

\subsection{Application: Gradient Descent on a Least Squares Problem}
We used the proposed method in solving a least squares problem via gradient descent uaing AWS Lambda. The problem that we focus on in this subsection is the standard least squares problem:
\begin{align} \label{eq_ls_problem}
    \text{minimize } ||Ax-y||_2^2
\end{align}
where $x \in \mathbb{R}^{n\times r}$ is not necessarily a vector; it could be a matrix. Let $x^* \in \mathbb{R}^{n\times r}$ be the optimal solution to \eqref{eq_ls_problem}, and let $x^*_i$ represent the $i$'th column of $x^*$. Note that if $x$ is a matrix, then the problem \eqref{eq_ls_problem} is equivalent to solving $r$ different least-squares problems with the same $A$ matrix: minimize $||Ax_i-y_i||_2^2$. The solution to this problem is the same as the $i$'th column of $x^*$, that is, $x^*_i$. When $r$ is large (and if we assume that $A$ fits in the memory of each worker), one way to solve \eqref{eq_ls_problem} in a distributed setting is by assigning the problem of minimizing $||Ax_i-y_i||_2^2$ to worker $i$. This solution however would suffer from possible straggling serverless function, and assumes that $A$ is small enough to fit in the memory of a serverless function. Another solution, the one we experiment with in this subsection, is to use gradient descent as in this case we are no longer restricted by $A$ having to fit in the memory and our proposed method can be used to provide resilience towards stragglers for this solution. 
\begin{figure}[htbp]
  \centering
  \includegraphics[width=0.4\textwidth]{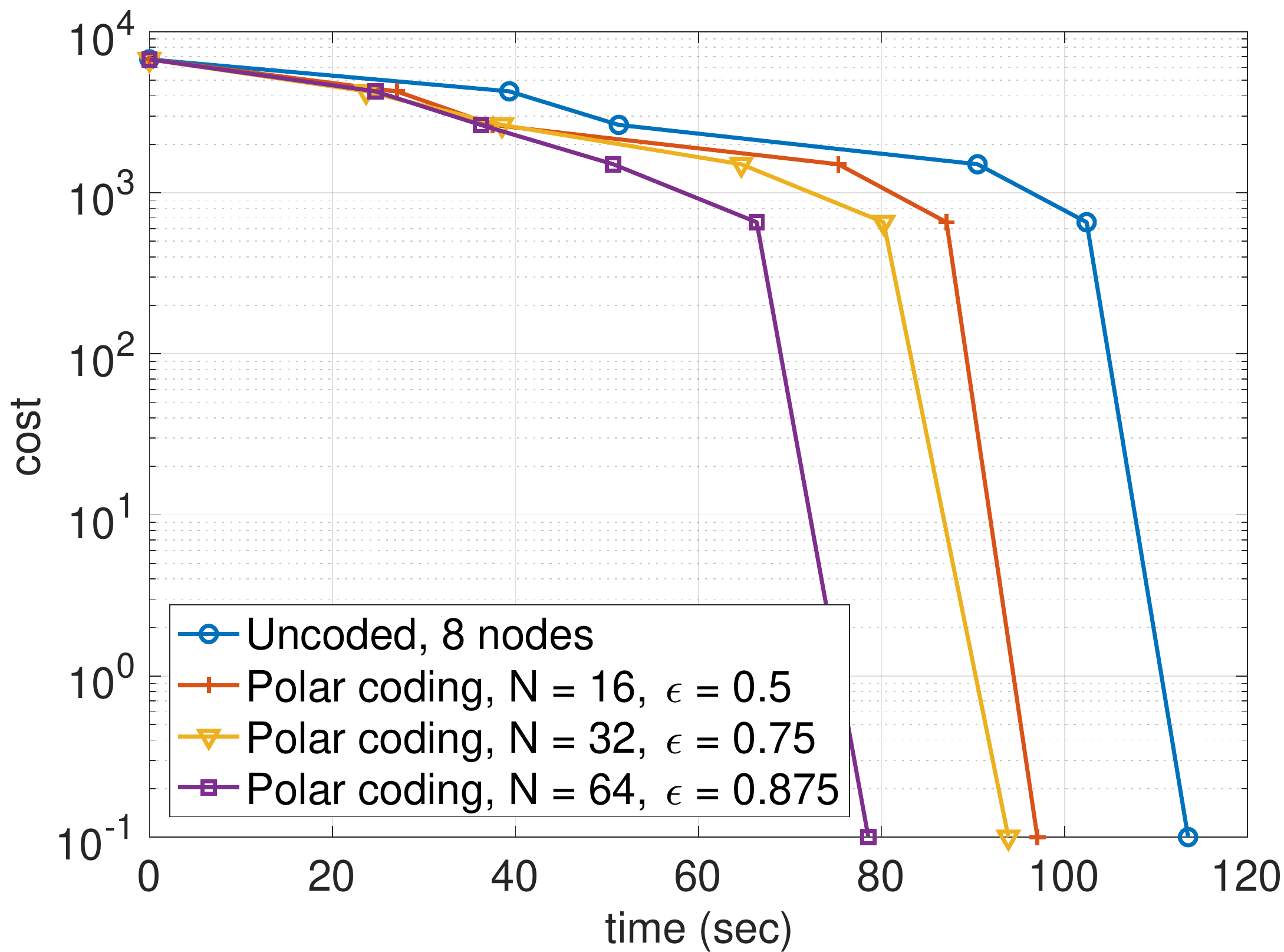}
  \caption{Cost vs time for the gradient descent example.}
  \label{plot_GD_8}
\end{figure}

The gradient descent update rule for this problem is:
\begin{align} \label{eq_update_rule}
    x_{t+1} = x_t - \mu (A^TAx_t - A^Ty),
\end{align}
where $\mu$ is the step size (or learning rate), and the subscript $t$ specifies the iteration number. 

We applied our proposed coded matrix multiplication scheme to solve a least squares problem \eqref{eq_ls_problem} using gradient descent where $A \in \mathbb{R}^{20000\times 4800}$, and $x \in \mathbb{R}^{4800\times 1000}$. Figure \ref{plot_GD_8} shows the downward-shifted cost (i.e., $l_2$-norm of the residuals $||Ax_t-y||_2$) as a function of time under different schemes. 
Before starting to update the gradients, we first compute and encode $A^TA$ offline. Then, in each iteration of the gradient descent, the master node decodes the downloaded outputs, updates $x_t$, sends the updated $x_t$ to AWS S3 and initializes the computation $A^TAx_t$. In the case of uncoded computation, we simply divide the multiplication task among $N (1-\epsilon)$ serverless functions, and whenever all of the $N (1-\epsilon)$ functions finish their computations, the outputs are downloaded to the master node, and there is no decoding. Then, master node computes and sends the updated $x_t$, and initializes the next iteration.

Note that in a given iteration, while computation with polar coding with rate $(1-\epsilon)$ waits for the first decodable set of outputs out of $N$ outputs, uncoded computation waits for all $N (1-\epsilon)$ nodes to finish computation. With this in mind, it follows that coded computation results in a trade-off between price and time, that is, by paying more, we can achieve a faster convergence time, as illustrated in Figure \ref{plot_GD_8}. Note that in coded computation we pay for $N$ serverless functions, and in uncoded computation we pay for $N (1-\epsilon)$ serverless functions. Using $\epsilon$ as a tuning parameter for redundancy, we achieve different convergence times.

\section{Conclusion}
We have proposed to use polar codes for distributed matrix multiplication in serverless computing, and discussed that the properties of polar codes are a good fit for straggler-resilient serverless computing. We have discussed the differences between serverless and server-based computing and addressed the differences and limitations of serverless computing in our proposed mechanism. We considered a centralized model with a master node and a large number of serverless functions. We implemented our proposed method using AWS Lambda and presented results showing its performance. We presented a sequential decoder algorithm that can encode and decode full-precision data so that our proposed framework can be used for computations on full-precision data. We discussed how to extend the proposed method to the two-dimensional case which corresponds to matrix multiplication where we code both matrices being multiplied. We have identified and illustrated with a numerical example a trade-off between the computation price and convergence time for the gradient descent algorithm applied to a least-squares problem.

\bibliographystyle{plain}
\bibliography{mert}

\end{document}